\documentclass[aps,pra,superscriptaddress,floats,twocolumn,longbibliography]{revtex4-1}
\usepackage{graphicx}
\usepackage{mathtools}
\usepackage{float}
\usepackage{mathrsfs}
\usepackage{graphicx}
\usepackage{natbib}
\usepackage{amsmath}
\usepackage{amssymb}
\usepackage{epsfig}
\usepackage{bm}
\usepackage{bbm}
\usepackage{subfigure}
\usepackage{dcolumn}
\usepackage{color}
\usepackage{epstopdf}
\usepackage{multirow}
\usepackage{booktabs}
\usepackage{times}
\usepackage{appendix}

\usepackage[plainpages=false,pdfpagelabels,colorlinks=true,linkcolor=red,urlcolor=blue,citecolor=blue,pdftitle={Title},pdfauthor={},pdfdisplaydoctitle=true,pdfduplex=DuplexFlipLongEdge]{hyperref}

\newcommand{\parc}[2]{\frac{\partial #1}{\partial #2}}

\def \be{\begin{equation}}
\def \ee{\end{equation}}
\def \bnm{\begin{enumerate}}
\def \en{\end{enumerate}}
\def \bea{\begin{align}}
\def \ea{\end{align}}
\def \bmtr{\begin{matrix}}
\def \emtr{\end{matrix}}
\def \bqt{\begin{quotation}\normalsize\sffamily}
\def \eq{\end{quotation}\large}
\def \bsm{\begin{normalsize}}
\def \es{\end{normalsize}}

\def \O{\Omega}
\def \e{{\epsilon}}

\def \D{{\Delta}}
\def \d{{\delta}}

\def \r{{\rho}}

\def \cL{{\mathcal{L}}}

\newcommand{\ket}[1]{\left\vert #1 \right\rangle}
\newcommand{\bra}[1]{\left\langle #1 \right\vert}
\newcommand{\ketbra}[2]{\ket{ #1}\bra{ #2} }

\newcommand{\tabincell}[2]{\begin{tabular}{@{}#1@{}}#2\end{tabular}}

\begin{document}


\title{Optimal quantum optical control of spin in diamond}
\author{Jiazhao Tian}
\affiliation{School of Physics, Huazhong University of Science and Technology, Wuhan 430074 P. R. China}
\affiliation{International Joint Laboratory on Quantum Sensing and Quantum
Metrology, Huazhong University of Science and Technology, Wuhan 430074,
China}
\author{Tianyi Du}
\affiliation{School of Physics, Huazhong University of Science and Technology, Wuhan 430074 P. R. China}
\author{Yu Liu}
\author{Haibin Liu}
\affiliation{School of Physics, Huazhong University of Science and Technology, Wuhan 430074 P. R. China}
\affiliation{International Joint Laboratory on Quantum Sensing and Quantum
Metrology, Huazhong University of Science and Technology, Wuhan 430074,
China}
\author{Fangzhou Jin}
\email{fzjin@hust.edu.cn}
\affiliation{School of Physics, Huazhong University of Science and Technology, Wuhan 430074 P. R. China}
\author{Ressa S. Said}
\affiliation{Institute for Quantum Optics and Center for Integrated Quantum Science and Technology, Ulm University, D-89081 Ulm, Germany}

\author{Jianming Cai}
\affiliation{School of Physics, Huazhong University of Science and Technology, Wuhan 430074 P. R. China}
\affiliation{International Joint Laboratory on Quantum Sensing and Quantum
Metrology, Huazhong University of Science and Technology, Wuhan 430074,
China}

\date{\today}

\begin{abstract}
The nitrogen-vacancy (NV) center spin represents an appealing candidate for quantum information processing. Besides the widely used microwave control, its coherent manipulation may also be achieved using laser as mediated by the excited energy levels. Nevertheless, the multiple levels of the excited state of NV center spin make the coherent transition process become complex and may affect the fidelity of coherent manipulation. Here, we adopt the strategy of optimal quantum control to accelerate coherent state transfer in the ground state manifold of NV center spin using laser. The results demonstrate improved performance in both the speed and the fidelity of coherent state transfer which will be useful for optical control of NV center spin in diamond.
\end{abstract}

\maketitle

\section{Introduction}\label{section1}

Nitrogen-vacancy (NV) color centers in diamond \cite{Gruber.S1997,Jelezko.PRL2004,Jelezko.PSSA2006,Dohertya.PR2013} have recently attracted increasing interest as an appealing solid state spin system for quantum information processing, including quantum computing and quantum sensing. The electronic spin associated with a single NV center demonstrates very long coherence times \cite{Balasubramanian.NM2009,Maurer.S2012}, the state of which can be efficiently initialised and readout using optical techniques \cite{Gruber.S1997}. Inspired by these extraordinary properties, much effort has been dedicated to use NV center spin as a building block for scalable room temperature quantum information processing \cite{Wrachtrup.J2006,Sar.N2012,Shi.PRL2010,Cai2013,Arroyo.NC2014,Barfuss2015,Shu.prl2018,Yu2018} and quantum sensing \cite{Balasubramanian.N2008,Maze.N2008,Grinolds.NP2013,Cai2014,Muller.NC2014,Sushkov2014,Shi.S2015,Rong.nc2018} as well as fundamental physics test \cite{Waldherr.PRL2011,Hensen.N2015,Jin.PRA2017}. In all of these applications, the efficient coherent manipulation of NV center spin is an indispensable ingredient, which is usually achieved by conventional electron spin resonance (ESR) techniques using microwave  \cite{Jelezko2004,Rong.PRL2014,Rong.nc2015} .

Recently, an all-optical control protocol for the electric spin of NV center has been demonstrated \cite{Christopher.PNAS2013,Christopher.np2016,Brian.NP2016}, which provides an efficient way for coherent manipulation of NV center spin and facilitates its applications in quantum optics \cite{Chu.Oxford2015} and magnetic resonance spectroscopy \cite{Wang_2014}. As compared with microwave control, all-optical spin control allows addressing of individual NV center spins and a faster speed of coherent spin manipulation. The all-optical quantum control of NV center spin is essentially based on coherent population transfer (CPT) \cite{Bergmann.RMP1998} and stimulated Raman adiabatic passage (STIRAP) techniques \cite{Vitanov.RMP2017} which exploits a $\Lambda$ system of two ground-state spin sublevels with an excited-state \cite{Christopher.PNAS2013,Hilser.PRB2012}. Therefore, the complicated energy levels of the excited-state of NV center spin would affect the achievable fidelity of optical manipulation. To achieve a fast optical control of NV center spin with a high fidelity, here we consider the optimisation of optical control by designing optimal laser driving fields.

Optimal control theory (OCT) \cite{Alessandro.Book2007,Glaser.EPJD2015} exploits numerical optimisation methods \cite{Fortunato.JCP2002,Caneva.PRA2011,Machnes.PRA2011,Ciaramella.JSC2015} to find the best control fields that steer the dynamics of a system towards the desired goal. Quantum optimal control can eliminate the effects of the rotating wave approximation and relax the adiabatic requirement, which has been successfully applied in the case of few-body systems \cite{Ryan.PRL2010,Machnes.PRL2010,Dolde.nc2014,Scheuer.njp2014, Geng.prl2017} as well as in ensembles \cite{Khaneja.JMR2005,Tosner.JMR2009,Li.prl2017} and correlated many-body quantum systems \cite{Doria.PRL2011,Frank.SR2016}. In this paper, we use optimal control theory to design all-optical control of the electric spin of NV center in diamond with high performance. We consider the Hamiltonian of the ground state spin triplet in the basis $\{\ket{0},\ket{+1},\ket{-1} \} $ and the excited state in the basis of spin-orbit states with full $C_{3v}$ symmetry $\{A_2,A_1,E_X,E_Y,E_1,E_2\}$, and a metastable intermediate state, see Fig.\ref{fig1}. We also take into account the effect of dissipation using a quantum master equation to describe the system's dynamics. Our goal is to optimise the state transfer efficiency from the ground state $\ket{-1}$ to $\ket{+1}$ via the excited state $A_2$ by optical control. The transfer efficiency as quantified by the fidelity is dependent on the power and shape of the laser driving fields. We adopt optimal control theory to speed up the coherent state transfer process with an improved transfer fidelity. Our result is expected to find applications in the further development of all-optical quantum control for NV center spin in diamond. We note that direct state transfer between the states $\ket{-1}$ and $\ket{+1}$ may also be achieved using strain \cite{Barfuss2015}. The present result provides an efficient way to achieve such a goal of NV spin coherent control.

The structure of the paper is the following. In Sec.\ref{section2} we elucidate the energy levels of the NV center spin and the description for the system's dynamics including laser driving and dissipation. In Sec.\ref{section3}, we investigate the performance of coherent state transfer using the scheme of STIRAP in the four-level and ten-level model  with and without dissipation. The optimal control theory and optimisation results are presented in Secs.\ref{section4}. Finally, in Sec.\ref{section5} we make a summary of our work.

\section{Model}\label{section2}

In this section, we first provide the details on the energy levels of the NV center spin and the description for the system's dynamics including laser driving and dissipation, which provides a starting point for our analysis of optimal control.

\subsection{Energy levels of NV center spin}

\begin{figure}[t]
\hspace{-0.1cm}
\includegraphics[width=1\columnwidth]{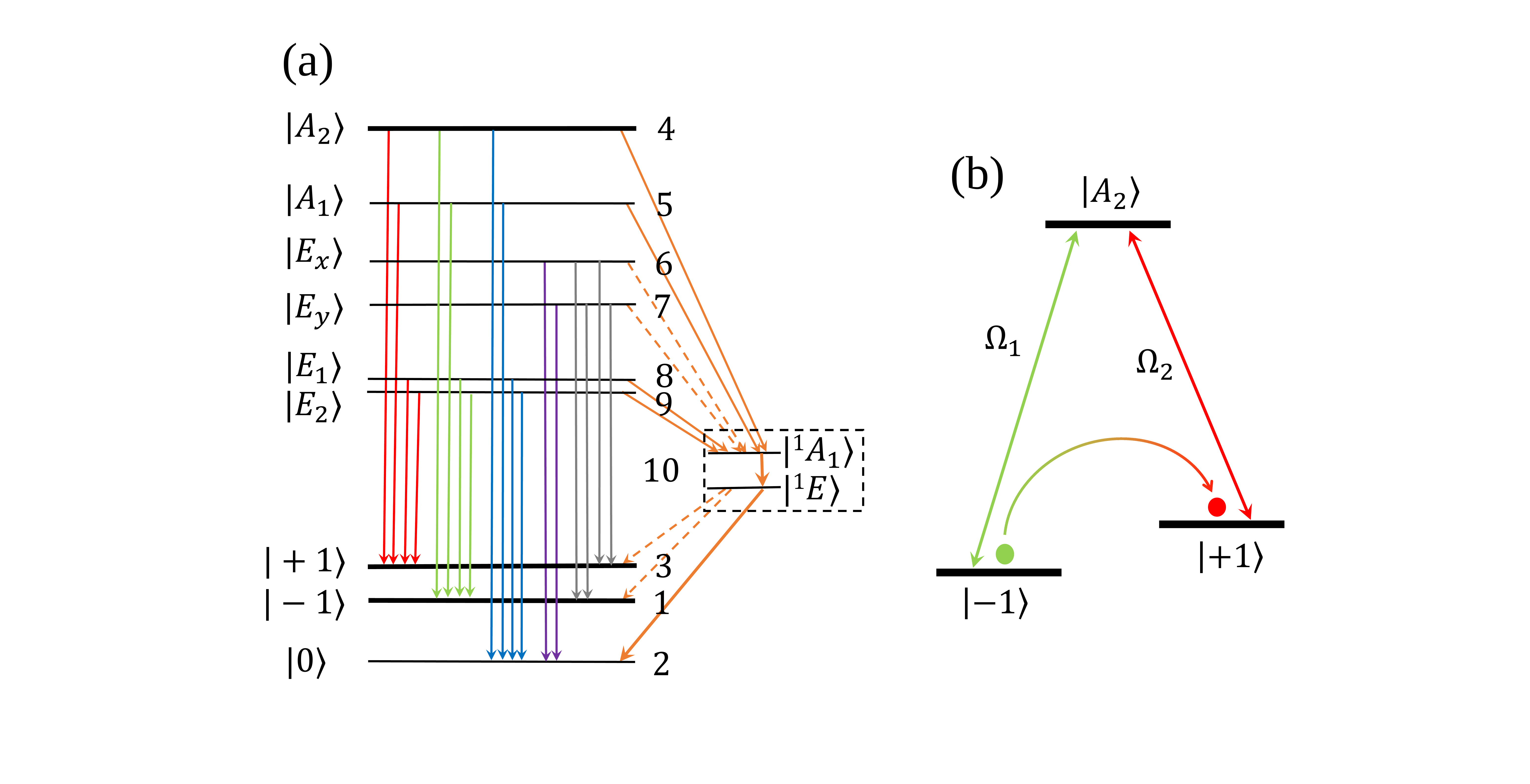}\\
  \caption{(Color online) (a) The sketch of the energy level structure of a single NV center in diamond under a magnetic field $B$, which consists of the upper six excited states  and the lower three ground states which are denoted by numbers according to the order in the Hamiltonian. The metastable states $\ket{^1A_1}$ and $\ket{^1E}$ are simply represented by $\ket{10}$. The arrows describe all possible decay channels between the excited states and the ground states. Solid arrows indicate relatively strong transitions, whereas dashed arrows indicate weak transitions which are neglected in our numerical calculation. (b) A simplified energy level scheme for stimulated Raman transition from the NV center energy level structure. We choose the excited state $\ket{A_{2}}$ and the ground states $\mid \pm1\rangle$ as a $\Lambda$ system. The lasers couple the ground states $\mid \pm1\rangle$ to the excited state $\ket{A_{2}}$ with the amplitudes $\Omega_{1}$ and $\Omega_{2}$ respectively.}\label{fig1}
\end{figure}

As a well studied atomic spin system, the negatively-charged NV center has six electrons, five of which are from the nitrogen and an extra electron is located at the vacancy site forming a spin $S=1$ pair. Because of the way they occupy the orbital states, the electronic structure of the NV center satisfies the $C_{3v}$ symmetry. The electron and orbital structures lead to a spin-1 triplet ground state manifold with a zero-field energy splitting of $D_{gs}\cong (2\pi)$ 2.88 GHz between its $m_s=\pm 1$ and $m_s=0$ sublevels. The Coulomb interaction results in an optical transition between the ground states and excited states with an energy gap $1.94$ eV. In the basis $\left\{\ket{-1},\ket{0},\ket{+1}\right\} \equiv \{m_s=-1, m_s=0, m_s=+1\} $, the Hamiltonian of the ground state spin triplet can be written as \cite{Hilser.PRB2012}
\be
H_{gs}=(D_{gs}-g_{gs}\mu_{B}B )|-1\rangle\langle -1|+(D_{gs}+g_{gs}\mu_{B}B )|+1\rangle \langle +1|,
\ee
 where $g_{gs}=2.01$ is the Land\'{e}-factor, $\mu_{B}$ is the Bohr magneton, and $B$ is the external magnetic field aligned with the NV axis. For simplicity, we set $\hbar=1$ in the whole text.
At low temperatures, taking into account the spin-spin interaction and spin-orbit interaction, the full Hamiltonian for the excited state manifold of the NV center spin can be written as the following $6\times6$ matrix in the basis of $\{A_2,A_1,E_X,E_Y,E_1,E_2\}$:
\cite{Hilser.PRB2012,Maze.njp2011,Chu.Oxford2015}:
\begin{equation}
H_{es}=
 \begin{bmatrix}
 H_{es}^{(I)} & 0 \\
  0 &  H_{es}^{(II)}
 \end{bmatrix} +E_g \mathrm{I}
\end{equation}
with
\begin{equation}
H_{es}^{(I)}=
 \begin{bmatrix}
 \Delta+2l_{z} & g_{es}\mu_{B}B \\
g_{es}\mu_{B}B &  -\Delta+2l_{z}
 \end{bmatrix},
\end{equation}
and
\begin{equation}
H_{es}^{(II)}=
 \begin{bmatrix}
-D_{es}+l_{z}& 0& 0 & \Delta^{''}\\
0 & -D_{es}+l_{z} &  i\Delta^{''} & 0\\
0 & -i\Delta^{''}& 0&-g_{es}\mu_{B}B\\
\Delta^{''} & 0 & -g_{es}\mu_{B}B & 0
 \end{bmatrix}
\end{equation}
where $D_{es}=(2\pi)1.42$ GHz, $\Delta=(2\pi)1.55$ GHz and $\Delta^{''}\simeq (2\pi)0.2$ GHz denotes the spin-spin interactions, $l_{z}=(2\pi)5.3$ GHz is the axial spin-orbit splitting, and $g_{es}\simeq 2.01$ is the Land\'{e}-factor of the excited state, $E_{g}$ denotes the energy gap between the lowest excited state and the ground states when there is no applied external magnetic field. For simplicity, we assume that the non-axial strain is zero, because it is negligible as compared with the other terms in the Hamiltonian.

\subsection{Coherent control of NV center spin using laser}

To achieve coherent control of the system, we consider applying a laser field with two frequencies $\omega_1$ and $\omega_2$ respectively, where $\omega_1$ ($\omega_2$) matches the energy gap between $\ket{A_2}$ and $\ket{-1}$ ($\ket{+1}$). The optical transitions between the ground states and the excited states arise from the electric dipole operator of two electrons as described by the Hamiltonian as $H_{dip}=H^{(1)}_{dip}\bigotimes\ \mathbbm{1}+\mathbbm{1}\bigotimes H^{(1)}_{dip}$, where $H^{(1)}_{dip}=e\bm{E}\cdot \hat{\bm{r}}$ is the single-particle electric dipole operator and $\hat{\bm{r}}=(\hat{x},\hat{y},\hat{z})$ denotes the electron position operator \cite{Hilser.PRB2012}. Here we assumed that the the laser field to be linearly polarized along the symmetry axis of the $e_x$ orbital so that the position operator is $\hat{\bm{r}}=\hat{x}$. Thus, the light-spin interaction can be described by the following matrix as
\be
V=
\left(
  \begin{array}{cc}
    0&v  \\
    v ^{\dagger}& 0 \\
  \end{array}
\right),
\ee
where
\be
v=\left(
 \begin{matrix}
   i\e_x&- i\e_x&0&0& -i\e_x&- i\e_x\\
    0&0&0&2\e_x&0&0\\
     -i\e_x&- i\e_x&0&0& i\e_x&- i\e_x\\
 \end{matrix}
    \right), \label{eq:transition_matrix}
 \ee
with
\be
\e_x=\frac{\bra{e_x}\hat{x}\ket{a}}{4}e E_x.
\ee
Here $\ket{a}$ represents the $a$ orbital state. We assume that $E_x=E_1 \cos(\omega_1 t)+E_2 \cos(\omega_2 t)$ is real as it represents a linearly polarized laser field, where $E_1$ and $E_2$ are the amplitudes with $\omega_1$ and $\omega_2$ the frequencies of the laser field. The energy gap between the ground states and the excited states is $E_g=1.94eV\approx 470$ THz. In the interaction with respect to $H_{E_g}=E_g\sum_{k=4}^{9}\ketbra{k}{k}$, the effective Hamiltonian can be written as follows
\be\label{eq:IntH}
H_I=e^{iH_{E_g}t}(H_{tot}-H_{E_g})e^{-iH_{E_g}t},
\ee
where $H_{tot}=H_{gs}+H_{es}+V$ is the total Hamiltonian. For the effective Hamiltonian $H_I$, the counter-rotating terms with frequencies $\omega_1+E_g$ and $\omega_2+E_g$ can be neglected. By defining the detuning $\delta_1=\omega_1-E_g$ and $\delta_2=\omega_2-E_g$, the transition matrix element $\e_x$ in the transition matrix $v$ (see Eq.\ref{eq:transition_matrix}) becomes
\be
\e_x'=\frac{\bra{e_x}\hat{x}\ket{a}}{4}e \left[\frac{E_1}{2} \cos\left(\delta_1 t\right)+\frac{E_2}{2} \cos\left(\delta_2 t\right)\right],
\ee
which can be simplified as
\be
\label{epsx}
\e_x'=\Omega_1\cos(\d_1t)+\Omega_2\cos(\d_2t)
\ee
with $\Omega_1=\bra{e_x}\hat{x}\ket{a}e E_1/8$ and $\Omega_2=\bra{e_x}\hat{x}\ket{a}e E_2/8$.
In addition, we consider the metastable state $\ket{10}$ (see Fig.\ref{fig1}) in the Hamiltonian as well, therefore the Hamiltonian that we use in the numerical simulation and optimisation has a total dimension of ten.

\subsection{Dissipative quantum master equation}

In this section, we will proceed to provide a formalism of quantum master equation to describe the system's dynamics. In order to take into account the influence of dissipation, we adopt a Lindblad form of quantum master equation as follows
\begin{eqnarray}
 \dot{\rho}(t)&&=-i[H,\rho(t)]\label{eq:qme}\\\nonumber
 &&+\sum \frac{\Gamma_{k\rightarrow l}}{2} \left[O_{k\rightarrow l}\rho(t)O_{k\rightarrow l} ^{\dag}-O_{k\rightarrow l}^{\dag}O_{k\rightarrow l}\rho(t) +h.c.\right]
\end{eqnarray}
The jump operator $O_{k\rightarrow l}$ represents the decay process from the $k$-th energy level to the $l$-th energy level at a rate $\Gamma_{k\rightarrow l}$. In our model, the dissipations from the excited states to the ground states are shown in Fig.\ref{fig1}(a). The decay rates between different energy levels are listed in Table \ref{table}. Note that the dephasing and relaxation time between the ground states in NV center are both longer than the lifetime of the excited states, hence the dephasing and relaxation from the ground states $\ket{\pm1}$ to $\ket{0}$ can be neglected.

\begin{table}[h]
\center
\caption{The decay rates between different energy levels of NV center spin, with the parameters adopted from Ref. \cite{Manson.PRB2006,Christopher.np2016}.}
	\label{table}
\begin{tabular}{p{5cm}<{\centering}|p{3cm}<{\centering}}
\toprule[1pt]
   decay rate $\Gamma_{k\rightarrow l}$& value  \\
   \hline
  $\ket{A_2},\ket{A_1},\ket{E_1},\ket{E_2}\rightarrow \ket{+1}$   & {$1/24 ~\text{ns}^{-1}$} \\
   \hline
  $\ket{A_2},\ket{A_1},\ket{E_1},\ket{E_2}\rightarrow \ket{-1}$  &{$1/31 ~\text{ns}^{-1}$}\\
    \hline
  $\ket{A_2},\ket{A_1},\ket{E_1},\ket{E_2}\rightarrow \ket{0}$  &{$1/104 ~\text{ns}^{-1}$}\\
    \hline
  $\ket{A_2},\ket{A_1},\ket{E_1},\ket{E_2}\rightarrow \ket{10}$  &{$1/33 ~\text{ns}^{-1}$}\\
    \hline
  $\ket{E_x},\ket{E_y}\rightarrow \ket{0}$& {$1/13~\text{ns}^{-1}$}\\
  \hline
  $\ket{E_x},\ket{E_y}\rightarrow \ket{+1},\ket{-1}$& {$1/666~ \text{ns}^{-1}$}\\
  \hline
   $\ket{E_x},\ket{E_y}\rightarrow \ket{10}$& {$0$}\\
  \hline
  $\ket{10}\rightarrow\ket{0}$&$1/303 ~\text{ns}^{-1}$\\
  \hline
  $\ket{10}\rightarrow\ket{\pm1}$&$0$\\
  \bottomrule[1pt]
\end{tabular}
\end{table}

\section{Performance of STIRAP with multiple excited levels}\label{section3}

In this section, we first investigate the performance of coherent state transfer using the scheme of STIRAP in the multi-level configuration of NV center spin, and demonstrate the influence of multiple excited levels on the fidelity of coherent state transfer.

\subsection{STIRAP control of NV center spin}

\begin{figure}[b]
\centering
\includegraphics[width=1\columnwidth]{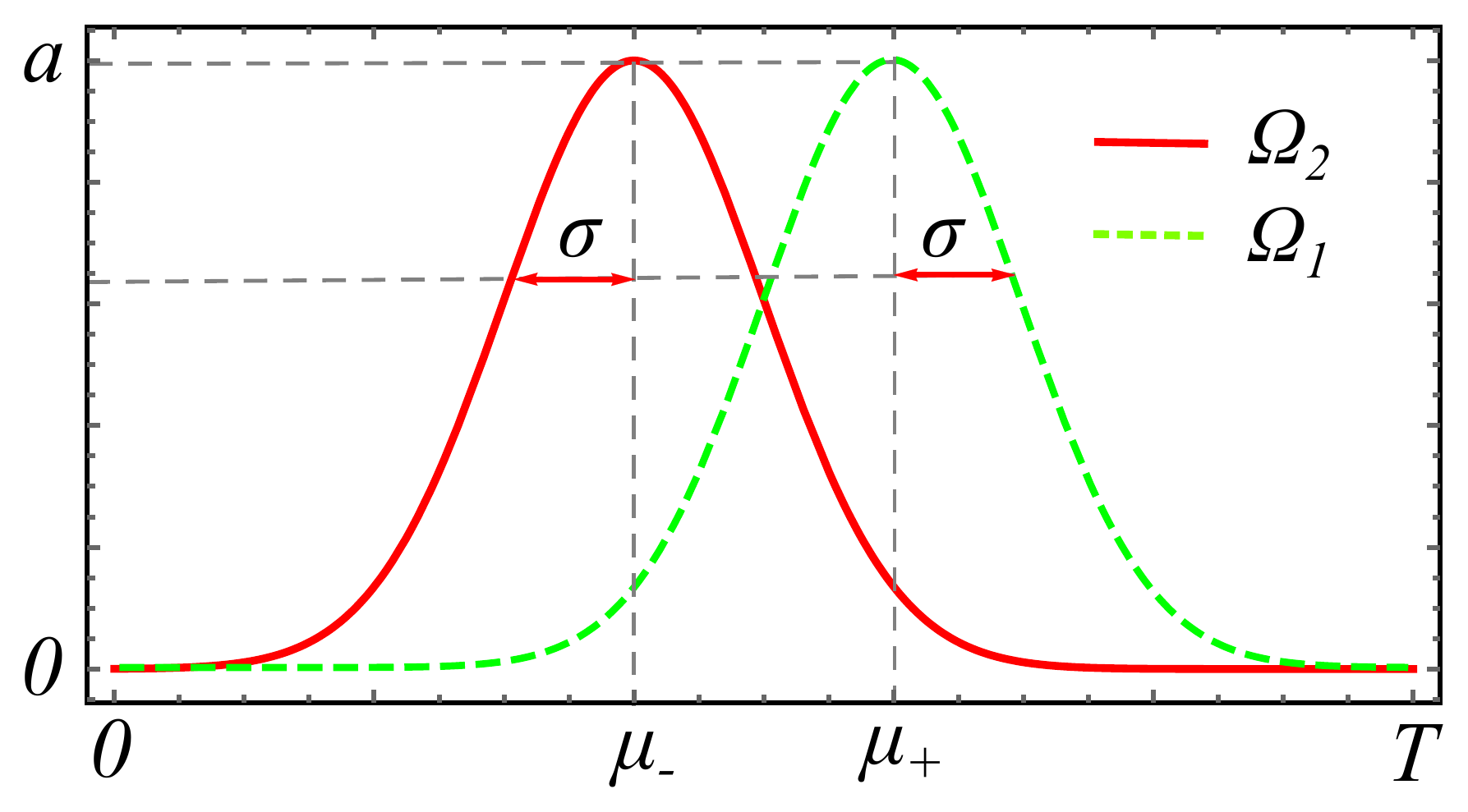}\\
\caption{(Color online) The Gaussian profile of the laser field amplitudes $\O_1$ (green dashed line) and $\O_2$ (red line) for the realisation of STIRAP, where $\Omega_1(t)$ and $\Omega_2(t)$ have the same maximum value $a$ (at time $\mu_+$ and $\mu_-$ respectively) and standard deviation $\sigma$.
}\label{Gaussianshape}
\end{figure}

All optical control of NV center spin makes use of a $\Lambda$-system  between the ground state $\ket{\pm 1}$ and the excited state $\ket{A_2}$ to implement coherent population trapping (CPT). The total Hamiltonian in the Hilbert subspace $\{\ket{-1},\ket{A_2},\ket{+1}\}$ is $H_s=H_0+V$ where
\be\label{simple3level}
H_0=\omega_a\ketbra{A_2}{A_2}+\omega_b\ketbra{-1}{-1}+ \omega_c\ketbra{+1}{+1},
\ee
with the parameters $\omega_a=\D+2l_z$, $\omega_b=D_{gs}-g_{gs}\mu_{B}B$, and $\omega_c=D_{gs}+g_{gs}\mu_{B}B$, and
\be
V=-i\e_x'\ketbra{A_2}{-1}+i\e_x'\ketbra{A_2}{+1}+h.c.,
\ee
where $\e_x'$ is expressed in Eq.\ref{epsx}.
For simplicity, we assume the conditions that $\Omega_1(t)\ll \omega_a-\omega_b=\delta_1$, $\Omega_2(t)\ll \omega_a-\omega_c=\delta_2$.
In the interaction picture with respect to $H_0$, the instantaneous eigenstates of the system's Hamiltonian $H_s^I$ can be written as follows
\begin{eqnarray}
\ket{D_0}&=& \frac{1}{\Omega(t)}[\Omega_2(t)\ket{-1}+\Omega_1(t)\ket{+1}]\\ \nonumber
\ket{D_{\pm}}&=&\frac{1}{\sqrt{2}}\left\{\mp i\ket{A_2}- \frac{1}{\Omega(t)}[\Omega_1(t)\ket{-1}-\Omega_2(t)\ket{+1}]\right\},
\end{eqnarray}
with the corresponding eigenvalues $\lambda_0=0$ and $\lambda_{\pm}=\pm \Omega(t)/2$, where $\Omega^2(t)={\Omega_1^2(t)+\Omega_2^2(t)}$. Starting from the initial state $\ket{-1}$ (i.e. $\ket{D_0(t=0)}$), we adiabatically tune the amplitude $\Omega_1$ from zero to a maximum value $a$, while tuning $\Omega_2$ from it's maximum $a$ to zero, see Fig.\ref{Gaussianshape}. Ideally, the system would end up in the target state $\ket{+1}$ if the change of the parameters satisfies the adiabatic condition as $\vert \langle{D_{\pm}}\vert{\dot D_0}\rangle/ \Omega(t)\vert \ll1$.
In our numerical simulation, the amplitudes of the laser field are chosen to be Gaussian functions of time, as shown in Fig.\ref{Gaussianshape}, namely
\begin{eqnarray}
\label{Gaussian_Omega1}
\O_1(t)&=&ae^{-(t-\mu_+)^2/{2\sigma^2}},\\
\label{Gaussian_Omega2}
\O_2(t)&=&ae^{-(t-\mu_-)^2/2\sigma^2}.
\end{eqnarray}
where the parameters $a$ and $\sigma$ are the maximum value and the standard deviation respectively. As an example, we take the parameters $\sigma=T/10$ and $\mu_{\pm}=T/2\pm\sigma$ for numerical simulation, where $T$ is the total evolution time.

\subsection{Influence of multiple excited levels on STIRAP}

\begin{figure}[t]
\hspace{0cm}
\includegraphics[width=1.05\columnwidth]{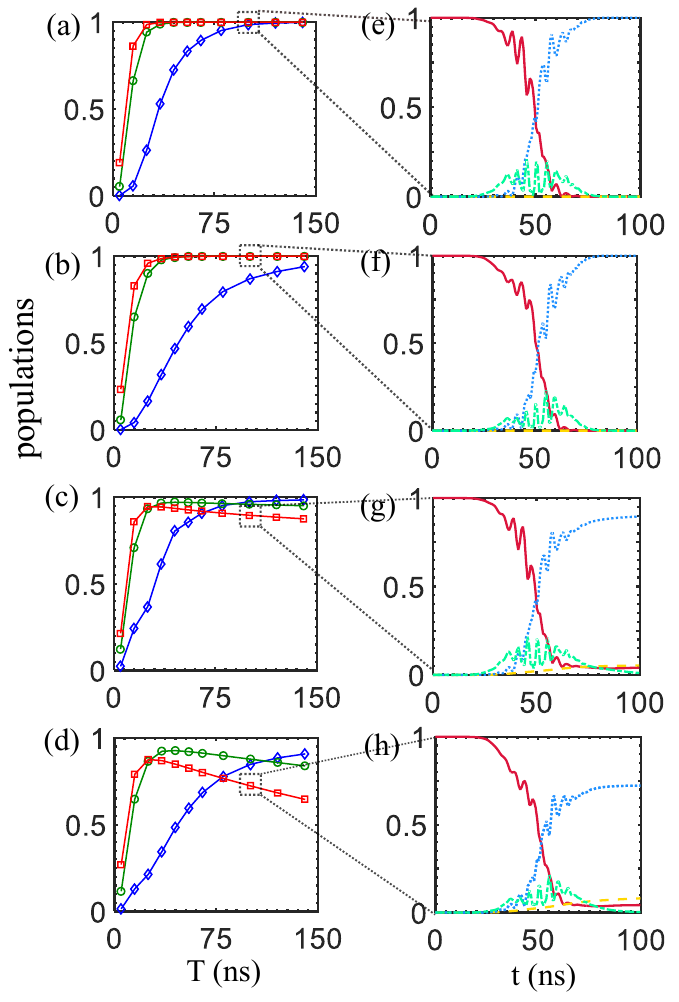}\\
 \caption{(Color online) (a-d) The final population of the target state $\ket{+1}$ in different situations as a function of the total evolution time $T$ of the STIRAP process. The results are shown for: (a) The simplified 4-level model which consists of $\{\ket{0},\ket{+1},\ket{-1},\ket{A_2} \}$ without dissipation; (b) The 10-level model which consists of $\{\ket{-1},\ket{0},\ket{+1},\ket{A_2},\ket{A_1},\ket{E_X},\ket{E_Y},\ket{E_1},\ket{E_2},\ket{10} \}$ without dissipation; (c) The 4-level model with dissipations from $\ket{A_2}$ to $\ket{\pm 1}$ and $\ket{0}$; (d) The 10-level model with all dissipations showed in Fig.\ref{fig1}.  The parameters of the control laser field are: $a=1$ GHz (blue diamond), $3$ GHz (green circle), $5$ GHz (red square), $\sigma=T/10$ and $\mu_{\pm}=T/2\pm\sigma$. The decay rates are presented in Table \ref{table}. (e-h) The dynamic evolution of the population of the state $\ket{-1}$ (red solid line), $\ket{+1}$ (blue dotted line), $\ket{A_2}$ (green dashed-dotted line), $\ket{0}$ (yellow dashed line) from the corresponding models with the parameters $a=5$ GHz and $T=100$ ns.
 }\label{fig3}
\end{figure}
In order to characterize the performance of STIRAP of NV center spin, in particular to investigate the influence of multiple excited levels and dissipation, we perform numerical simulation with the Hamiltonian
\be
H=H_{gs}\otimes I +I\otimes H_{es}+V',
\ee
where
\be
V'=
\left(
  \begin{array}{cc}
    0&v\\
    v ^{\dagger}& 0 \\
  \end{array}
\right),
\ee
and
\be
v=\left(
 \begin{matrix}
   i\e'_x&- i\e'_x&0&0& -i\e'_x&- i\e'_x\\
    0&0&0&2\e'_x&0&0\\
     -i\e'_x&- i\e'_x&0&0& i\e'_x&- i\e'_x\\
 \end{matrix}
    \right). \label{eq:transition_matrix}
 \ee
We make an approximation that the continuous field is approximated by successive small time intervals $\D t$, during which the amplitudes of the field are assumed to be constant.  The density operator $\rho(t)$ at time $T=N\D t$ can be written as
 \be
 \rho(T)=e^{\hat{\mathcal{L}}_N \Delta t}\cdots e^{\hat{\mathcal{L}}_1 \Delta t}\rho_0,
 \ee
 where $\hat{\mathcal{L}}_j=-i\hat{H}_j+\hat{\Gamma}$ is the Liouville superoperator, $\hat{H}_j$ is the Hamiltonian superoperator and $\hat{\Gamma}$ is the relaxation superoperator \cite{Khaneja.JMR2005}.
The population of the state $\ket{k}$ at time $T$ is given by $P_{\ket{k}}(T)=\bra{k}\rho(T)\ket{k}$.
 The numerical simulation converges when $\D t$ is sufficiently small. We use the Runge-Kutta method to test the convergence of our numerical simulation. To tune the laser field continuously, we set $\Delta t=0.005$ ns in the Runge-Kutta method, the convergence of which has been numerically verified. In the following numerical simulation, the time step $\Delta t$ is set to be $0.005$ ns as well.

We compare the following four different situations as: (1) The 4-level model consists of $\{\ket{0},\ket{+1},\ket{-1},\ket{A_2} \}$ without dissipation;  (2) The 10-level model consists of $\{\ket{-1},\ket{0},\ket{+1},\ket{A_2},\ket{A_1},\ket{E_X},\ket{E_Y},\ket{E_1},\ket{E_2},\ket{10} \}$ without dissipation ; (3) The 4-level model with dissipations from $\ket{A_2}$ to $\ket{\pm 1}$ and $\ket{0}$;  (4) The 10-level model with all dissipations showed in Fig.\ref{fig1}. We choose three different control laser fields with the maximum amplitude $a=1,3,5$ GHz, and the results are shown in Fig.\ref{fig3}. Here we consider the ground state $\ket {-1}$ as the initial state of NV center spin. We will compare the fidelity of the target state $\ket {+1}$ after a fixed time with the above different models.

In Fig.\ref{fig3}, we first show the fidelity of coherent state transfer (i.e. the final population of the state $\ket{+1}$ after the STIRAP process) for different laser amplitudes $a$ and total evolution time $T$. It can be seen that a sufficient long total evolution time is necessary to ensure a high transfer fidelity by satisfying the adiabatic condition. The state transfer fidelity significantly decreases when the total evolution becomes shorter, e.g. than 50 ns. On the other hand, in the adiabatic regime, a relatively larger laser field amplitude would improve the state transfer fidelity. The comparison between Fig.\ref{fig3}(a) and Fig.\ref{fig3}(c) [Fig.\ref{fig3}(b) {\it vs.} Fig.\ref{fig3}(d)] shows that the involved extra excited levels (which are absent in the simplified 4-level model) lead to a worse performance. The results demonstrate that the multiple excited states apart from $\ket{A_2}$ would hinder the efficiency of coherent state transfer, and need to be taken into account in the analysis.
Comparing Fig.\ref{fig3}(b) with Fig.\ref{fig3}(d), it can be seen that the dissipation will degrade the performance of coherent state transfer. This fact implies that it is beneficial to accelerate the speed of state transfer. It is possible to resort to a large power of laser, which nevertheless will be in contradiction to the adiabatic requirement and thus cause more severe leakage to the excited state $\ket {A_2}$. In Fig.\ref{fig3}(e-h), we plot the detailed time dynamics of the population of the state $\ket{-1}$ (red solid line), $\ket{+1}$ (blue dotted line), $\ket{A_2}$ (green dashed-dotted line), $\ket{0}$ (yellow dashed line) for the parameters $a=5$ GHz and $T=100$ ns. The final populations on $\ket {+1}$ are $0.999992$, $0.999958$, $0.895$ and $0.722$ from top to bottom, respectively.
In the appendix \ref{Morecharacters}, we calculate the performance of STIRAP process for other more different parameters. It shows that too weaker or stronger pulses will result in a worse performance.
These results clearly demonstrate that the complicated energy levels of the excited-state of NV center spin and the dissipation affect the performance of STIRAP. To achieve an optically controlled coherent state transfer of NV center spin with a high fidelity in a short time, we will proceed to consider the optimisation of optical control by designing  optimal laser driving fields in the following section.

\section{optimisation of optical NV spin control}\label{section4}

In this section, we adopt optimal control theory to improve the efficiency of coherent state transfer of NV center spin using shaped laser pulse. We use both the GRAPE method and the Nelder-Mead method to perform optimisation.  In the following we first introduce the principle of the GRAPE method and the detailed formalism for the present system. We then proceed to illustrate four types of optimisation methods that we use. Finally, we investigate in detail the obtained optimal results in terms of coherent state transfer fidelity, required laser power and robustness.

\subsection{GRAPE method}
GRAPE is an efficient method \cite{Khaneja.JMR2005} for the engineering of pulse sequences in order to achieve optimal dynamical performance, e.g. state transfer efficiency and quantum gate fidelity. In the present scenario, the total Hamiltonian of the system can be divided into two parts: the time-independent Hamiltonian $H_0$ and the time-dependent Hamiltonian $H_c$ that is dependent on a set of time-dependent parameters $u_k(t)$. The total Hamiltonian is represented as follows
\be
H(t)=H_0+H_c\left[u_1(t),\cdots ,u_k(t),t\right].
\ee
The total evolution time $T$ is divided into a sequence of small time segments $\Delta t$, and the parameter $u_k$ is represented as $u_k(1),\cdots,u_k(j),\cdots,u_k(N)$, where  $N=T/\Delta t$ is the number of time segments. The value $u_k(j)$ is considered to be a constant during the corresponding time interval $[(j-1)\Delta t, j\Delta t]$. We denote $\phi$ as the target function to be maximised, thus the gradient-based iteration process is $
u_k(j)=u_k(j)+\epsilon \cdot \partial \phi/\partial u_k(j)$, where $\epsilon$ is an adjustable parameter to guarantee the convergency.

The target function consists of three parts as
\be\label{targetfunction}
\phi =p_3+\lambda\bar p_4+\lambda_E E.
\ee
where $p_3$ is the final population of the target state $\ket{+1}$, $\bar p_4$ is the average population on $\ket{A_2}$, and $E$ is the total power of the laser field $E=\sum_{j=1}^N\left[ \Omega^2_1(j)+\Omega^2_2(j) \right]$. $\lambda$ and $\lambda_E$ are the weight factors of $\bar p_4$ and  $E$ respectively. To maximize $\phi$ with negative values of $\lambda$ and $\lambda_E$ indicates searching for the highest transition rate to state $\ket{+1}$ while keeping the population on $\ket{A_2}$ and the total power of the laser field under certain constraints during the evolution process. The detailed derivation of $\phi$ with respect to the control parameter $u_k(j)$ is presented in the appendix \ref{Numerical details}.

\begin{table}[t]
\hspace{-0.5cm}
\caption{Four types of optimisation methods}
\label{table1}
\begin{tabular}{c|c|c|c}
\toprule[1pt]
Name & Method & Initial value & Parameters   \\
\hline
\multirow{5}{*}{Adiabatic-NM} & \multirow{5}{*}{Nelder-Mead} & \multirow{7}{*}{\tabincell{c}{Gaussian functions \\$a \in [0, 3]~ \mbox{GHz}$ \\ $\mu \in [\mbox{T}/4, 3\mbox{T}/4]$ \\ $\sigma \in [\mbox{T}/20, 3\mbox{T}/20]$} }&\multirow{5}{*}{ $a,\mu,\sigma$} \\
    & & &  \\
    & & &  \\
    & & &  \\
 \cline{1-2}  \cline{4-4}
 \multirow{3}{*}{Adiabatic-G}&\multirow{3}{*}{GRAPE}& &\multirow{3}{*}{$\O_1(j)$, $\O_2(j)$}\\
    & & &  \\
    & & &  \\
   \cline{1-4}
   \multirow{4}{*}{Rabi-resonant}&\multirow{4}{*}{GRAPE} & \multirow{4}{*}{\tabincell{c}{Constant functions\\
   $a\in [0, 3] ~\mbox{GHz}$ \\$\D=0$  }}&\multirow{4}{*}{$\O_1(j)$, $\O_2(j)$} \\
    & & &  \\
    & & &  \\
    & & &  \\
   \cline{1-4}
  \multirow{4}{*}{Rabi-detuning}&\multirow{4}{*}{GRAPE} & \multirow{4}{*}{\tabincell{c}{Constant functions\\
  $a\in [0, 3] ~\mbox{GHz}$ \\$\D\in [0, 3]~ \mbox{GHz}$ }}&\multirow{4}{*}{$\O_1(j),\O_2(j),\D$}  \\
   & & & \\
   & & &  \\
   & & &  \\
\bottomrule[1pt]
\end{tabular}
\label{table}
\end{table}

\subsection{Optimisation methods}

In order to avoid local optimal points, we choose the initial values for optimisation covering a relatively broad range in a random way. We consider four different types of optimisation method with different initial points and optimisation methods (see Table \ref{table}): \emph{(1) Adiabatic Nelder-Mead, (2) Adiabatic GRAPE, (3) Rabi resonant GRAPE} and \emph{(4) Rabi detuning GRAPE}. The Adiabatic Nelder-Mead method and the Adiabatic GRAPE method starts from the laser field of Gaussian functions (from STIRAP, see Fig.\ref{Gaussianshape}) as follows
\begin{eqnarray}
\O_1(t)&=&a\exp[-\frac{(t-\mu)^2}{2\sigma^2}],\\
 \O_2(t)&=&a\exp[-\frac{(t-(T-\mu))^2}{2\sigma^2}].
\end{eqnarray}
with randomly chosen parameters $a$,   $\mu $ and $\sigma$. The Adiabatic Nelder-Mead method optimises these parameters using Nelder-Mead algorithm, while the Adiabatic GRAPE method optimises $\Omega_1(j)$ and $\Omega_2(j)$ using GRAPE algorithm. The Rabi resonant and the Rabi detuning methods starts from the laser field with identical values of  $\Omega_1(j)$ and $\Omega_2(j)$ with a random amplitude $a$, and perform optimisation using GRAPE algorithm. The Rabi resonant GRAPE method uses resonant laser fields, and thus the parameter of detuning is $\Delta =0$. In contrast, the Rabi detuning GRAPE method also optimises the parameter of detuning $\Delta$ with a random initial value. Table \ref{table1} gives a summary of these four optimisation methods we use in this work. For the target function, we choose $\phi=p_3$ without involving the constraints on $\bar p_4$ and $E$, which leads to the highest coherent state transfer efficiency while the corresponding values of $\bar p_4$ and $E$ are verified to be within the reasonable limits. During the optimisation, we set the convergency criterion as $\phi(m+100)-\phi(m)<10^{-3}$, where $m$ is the number of iteration steps.

\subsection{Optimisation results}

\begin{figure}[t]
\hspace{0cm}
\includegraphics[width=1\columnwidth]{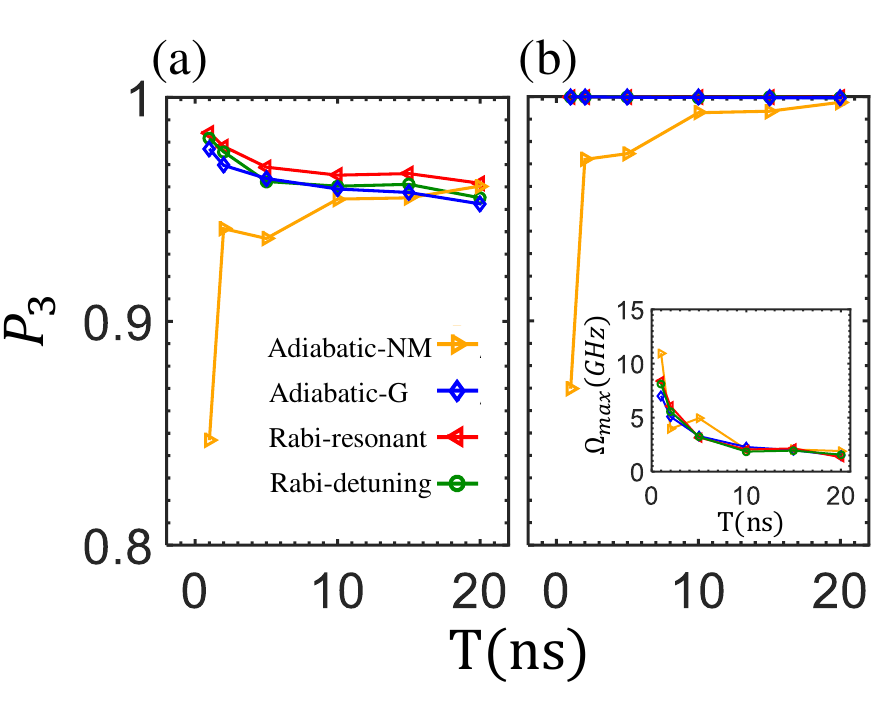}
\caption{(Color online) The optimal results from different evolution time $T$ using four optimisation methods (as listed in Table~\ref{table1}) of the 10-level model of NV center spin. Each data is based on $500$ random initial points.  (a) shows the final population $p_3$ of the target state $\ket{+1}$ under the influence of dissipations. (b) shows the final population $p_3$ of the target state $\ket{+1}$ without dissipation. The inset shows the corresponding maximum laser field amplitudes that achieve the optimal fidelity of coherent state transfer. The results from four types optimal methods are presented in different symbols: (1) adiabatic Nelder-Mead (yellow triangle), (2) adiabatic GRAPE (blue diamond), (3) Rabi resonant GRAPE (red triangle) and (4) Rabi detuning GRAPE (green circle).
}\label{fig4}
\end{figure}

The optimisation results for different values of total evolution time $T$ are summarised in Fig.\ref{fig4}. It can be seen from Fig.\ref{fig4}(a) that the fidelity of coherent state transfer from the adiabatic Nelder-Mead method, namely following an optimised STIRAP process, decreases significantly as the total evolution time becomes shorter. This result can be understood from the adiabatic condition underlying the STIRAP process, the breakdown of which for a short total evolution time would degrade its performance due to the excitation to the other eigenstates. For comparison, we find that the other three optimisation methods result in much better fidelities of coherent state transfer. The results demonstrate that the optimisation of optical control can significantly enhance the fidelity of coherent state transfer and accelerate the speed of optical coherent manipulation of NV center spin. Under optimal control, the coherent state transfer can be achieved with a high fidelity within the time on the order of nanosecond.

\begin{figure*}[t]
\hspace{0cm}
\includegraphics[width=2.0\columnwidth]{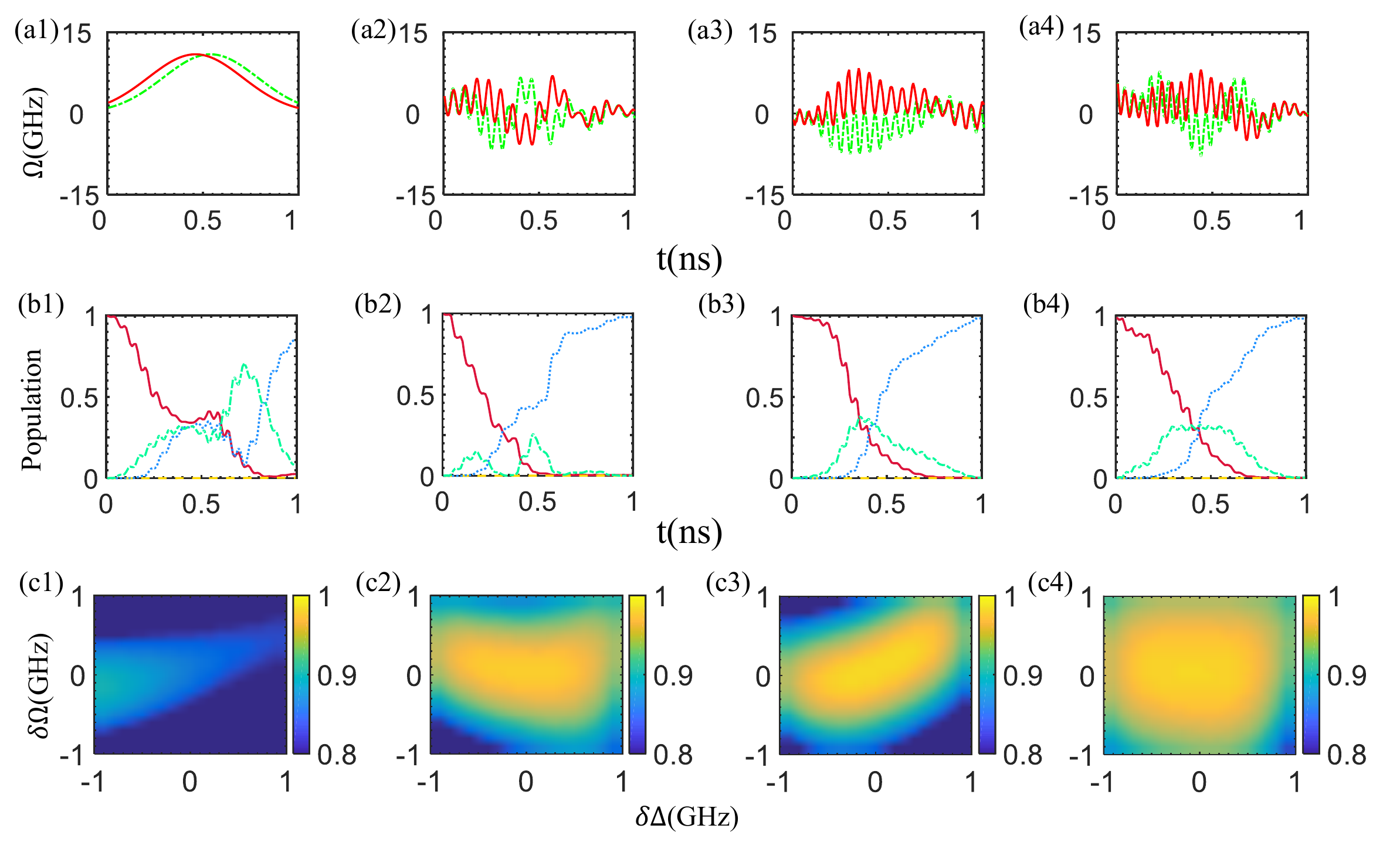}
\caption{(Color online) The details of the optimisation results. (a1-a4) show the amplitudes of the control laser field $\O_1(t)$ (green dashed line) and $\O_2(t)$ (red solid line) for the total evolution time $T=1$ ns (see Fig.\ref{fig4}) that are obtained by using the optimisation methods in Table~\ref{table1}.  (b1-b4) The dynamic evolution of the population of the state $\ket{-1}$ (red solid line), $\ket{0}$ (yellow dashed line), $\ket{+1}$ (blue dotted line) and $\ket{A_2}$ (green dashed-dotted line) for the corresponding four optimisation results of $T=1$ ns respectively. (c1-c4) The final population of the target state $\ket{+1}$ versus the amplitude systematic errors $\delta \O$ and the frequency detuning $\delta \Delta$ of the optimised control laser field. The four columns of figures correspond to the results using four optimisation methods: adiabatic Nelder-Mead (a1, b1, c1), adiabatic GRAPE (a2, b2, c2), Rabi resonant GRAPE (a3, b3, c3) and Rabi detuning GRAPE (a4, b4, c4), respectively. The optimal parameter of the detuning in the Rabi-detuning GRAPE method is $\D=9.4707$ GHz.
}\label{fig5}
\end{figure*}

To investigate the role of dissipation, we plot the fidelity of coherent state transfer in Fig.\ref{fig4}(b). It can be seen that coherent state transfer by optimal control can reach an almost unit fidelity if there is no dissipation.
As the total evolution time becomes shorter, the optimisation of $\Omega_1(j)$ and $\Omega_2(j)$ gives a better performance, because the influence of dissipation also becomes less pronounced. This is different from the result of an optimised STIRAP process (by the adiabatic Nelder-Mead method), where the performance is limited by the overall effect of the dissipation and the violation of adiabaticity. We note that the required maximum power of laser field is similar for four types of optimisation methods, as shown in the inset of Fig.\ref{fig4}(b), which is feasible in experiment.

In Fig.\ref{fig5}, we plot the details of the optimisation results for different optimisation methods.  In Fig.\ref{fig5}(a1-a4), we show the amplitudes $\O_{1}(t)$ and $\O_{2}(t)$ of the laser field that achieve the optimal coherent state transfer efficiency for a total evolution time $T=1$ ns. Fig.\ref{fig5}(b1-b4) show the corresponding dynamic evolution of the population of the state $\ket{-1}$, $\ket{0}$, $\ket{+1}$ and $\ket{A_2}$. The final populations on $\ket {+1}$ are  $0.8469$, $0.9770$, $0.9842$ and $0.9816$ from left to right, respectively.
It can be seen from Fig.\ref{fig5}(b1) that the state $\ket{A_2}$ is significantly populated, which suggests that the adiabatic condition as required by STIRAP process is not satisfied anymore when the total evolution time is not sufficiently long.

We further test the robustness of the obtained optimal control laser fields against the deviation in the amplitude of the laser field and the frequency detuning. In Fig.\ref{fig5}(c1-c4), we show the final population of the target state $\ket{+1}$ as a function of the laser field amplitude systematic errors \cite{Said2009} $\delta \O$ and the frequency detuning $\delta \Delta$ for the optimal results as obtained using different optimisation methods. It can be seen that the fidelity of coherent state transfer by the optimal control laser field is quite robust against the systematic errors in the control laser field, as shown in Fig.\ref{fig5}(c2-c4), demonstrating superior performance over the optimised STIRAP process [Fig.\ref{fig5}(a1)]. In particular, the optimal result, which is obtained by the Rabi detuning GRAPE method appears to be the best strategy in this regard, see  Fig.\ref{fig5}(c4). The precise timing of laser would also be important for the performance of optimal control. We remark that the optical control of the spin in NV center as demonstrated experimentally can reach a time scale of $1$ ns \cite{Bassett.S2014}. The laser pulse shape can be modulated with an even higher resolution \cite{Scheuer.njp2014} using a fast arbitrary wave generator. In the appendix \ref{STIRAP1ns}, we show the optimal result for the evolution time $T=1$ ns a time resolution of $0.05$ ns, which is feasible with  the state-of-the-art experiment capability. The final population of the target state $\ket{+1}$ can reach about 0.9765.
\section{Conclusion and discussion}\label{section5}
To summarize, we use the optimal control theory to improve the performance of all-optical control of the electric spin of NV center in diamond. We compare the fidelity of coherent state transfer of the simplified 4-level model with one single excited state and the 10-level model with the relevant multiple excited states under the influence of dissipation.  The results show that the complicated multiple energy levels of the excited-state of NV center spin and the dissipation affects the performance of the conventional STIRAP process, and thus put a constraint on the achievable fidelity and the speed of coherent state transfer. We adopt four different optimisation methods and obtain control laser fields that can achieve significantly improved fidelity of coherent state transfer. The speed of optical control of NV center spin is also enhanced, which can be realised on the order of nanoseconds. Moreover, we find that the performance of the optimal control laser fields is robust against the deviations in the amplitude and frequency of the laser field. The present results will facilitate the development of high-fidelity and fast-speed all-optical quantum control for NV center spin in diamond.

\section*{ACKNOWLEDGMENTS}
We thank Y. Chu and Prof. Martin B. Plenio for helpful discussion of optimal control theory and numerical simulation. This work is supported by the National Key R$\&$D Program of China (Grant No. 2018YFA0306600), the National Natural Science Foundation of China (11874024, 11574103, 11804110, 11690032), the Fundamental Research Funds for the Central Universities. RSS acknowledges support from ERC Synergy Grant BioQ and EU Project Asteriqs.

\setcounter{section}{0}
\setcounter{equation}{0}
\setcounter{figure}{0}
\setcounter{table}{0}

\renewcommand{\thesection}{A.\arabic{section}}
\renewcommand{\thesubsection}{\thesection.\arabic{subsection}}
\renewcommand{\theequation}{A\arabic{equation}}
\renewcommand{\thefigure}{A\arabic{figure}}

\section*{Appendix}

\section{Numerical details}\label{Numerical details}

\subsection{State evolution}
To calculate the evolution of the system's state with time numerically, we define  $\r ^{vec}$ as a vector constructed by rearranging the $N_E \times N_E$ density matrix $\rho$  into an $N_E^2 \times 1$ vector. The rearranging process is represented by
\be\label{reshpe1}
\r^{vec}=\mbox{reshape} (\r,N_E^2,1).
\ee
At time $t=j\D t$,the state of the system can be written as
\be
\r^{vec}_j=e^{\cL_j\Delta t}\r_{j-1}^{vec},
\ee
where the map $\cL_j$ is a $N_E^2\times N_E^2$ matrix, the $l$-th column of which can be calculated as
\be
\cL_j(:,l)=\mbox{reshape}(\mathcal{O}(\r^l),N_E^2,1), l=1,2,\cdots N_E^2.
\ee
with $\r^l=\ketbra{m}{n}$, here $m=\mbox{rem}(l,N_E)$ is the remainder of $l/N_E$, and $n=[l/N_E]+1$ where $[l/N_E]$ is the integer part of $l/N_E$, the map $\mathcal{O}(\r)=\dot{\rho}(t)$ is the right-hand side of quantum master equation in Eq.(\ref{eq:qme}). To avoid computing the time-independent part repetitively, the map $\cL$ is divided into two parts as follows
\be\label{Lj}
\cL_j=\cL_0+\cL _{\e_x'}\cdot \e_x'(j),
\ee
where $\cL_0$ is the time-independent part and  $\cL _{\e_x'}=\parc{\cL}{\e_x'} $ is the partial derivation of $\cL$ with respect to $\e_x'$, which is also time-independent.

\subsection{Formalism of GRAPE method}
In the following, we introduce the detailed algorithm for the GRAPE method of 10-level model. The laser field of the $j$-th time segment can be written as
\be
\e_x'(j)=\Omega_1(j)\cos\left[(\d_1+\Delta) t_j\right]+\Omega_2(j)\cos\left[(\d_2+\Delta) t_j\right],
\ee
where $\D$ is the time-independent detuning term. Therefore there are three sets of parameters for optimisation: $u_1(j)\equiv\Omega_1(j)$, $u_2(j)\equiv\Omega_2(j)$, and $u_3\equiv\Delta$. To get the term $\parc{\phi}{u_k}$ used in the iteration formula in GRAPE method, we calculate $p_3, \bar{p}_4$ and $E$ in equation (\ref{targetfunction}) respectively in below. The derivative of $p_3$ with respect to the control parameter $u_k(j)$ is
\be
\parc{p_3}{u_k(j)}=\mbox{Tr}[\hat{P}_{+1}\parc{\r_{N}}{u_k(j)}],
\ee
where $\hat{P}_{+1}=\ketbra{+1}{+1}$.
As shown in Eq.(\ref{reshpe1}) , the derivative of the density matrix $\parc{\r_{N}}{u_k(j)}$ can be mapped to a form of vector as
\be\label{plugfirst}
 \parc{\r_{N}}{u_k(j)}=\mbox{reshape}(\parc{\r^{vec}_{N}}{u_k(j)},N_E^2,1),
\ee
and vice versa,
\be
 \parc{\r^{vec}_{N}}{u_k(j)}=\mbox{reshape}(\parc{\r_{N}}{u_k(j)},N_E,N_E).
 \ee

To calculate the derivatives, we define the forward operator and backward operator as follows
\begin{eqnarray}
U_{forw}(j)&=&e^{\cL_j \D t}\cdots e^{\cL_{1}\D t },\\
U_{back}(j)&=&e^{\cL_N \D t}\cdots e^{\cL_{j+1} \D t}.
\end{eqnarray}
So that
\be\label{gradient}
\parc{\r^{vec}_{N}}{u_k(j)}=U_{back}(j) \parc{e^{\cL_j\Delta t}}{u_k(j)} \r^{vec}_{j-1}.
\ee
The derivative of the exponential term in the above equation is given by \cite{fisher2010optimal}
\be
\parc{e^{ \mathcal{L}_j\Delta t}}{u_k(j)}=\left[\int_0^1e^{s \mathcal{L}_j\Delta t}\left(\Delta t\parc{\mathcal{L}_j}{u_k(j)}\right)e^{-s \mathcal{L}_j\Delta t}ds \right] e^{\mathcal{L}_j\Delta t}.
\ee
For a small value of $\Delta t$, it can be approximated as
\be\label{pluglast}
\parc{e^{\mathcal{L}_j\Delta t}}{u_k(j)}=\Delta t\parc{\mathcal{L}_j}{u_k(j)}e^{ \mathcal{L}_j\Delta t},
\ee
which along with Eq. (\ref{Lj}) gives the simplified form of Eq.(\ref{gradient}) as follows
\be
\parc{\r^{vec}_{N}}{u_k(j)}=U_{back}(j)\cL_{\e_x'} \left(\D t\parc{\e_x'(j)}{u_k(j)}\right) U_{forw}(j)\r^{vec}_0.
\ee
For the third parameter $u_3\equiv\D$, the derivative is
\be
\parc{\r^{vec}_{N}}{u_3}=\sum_j^N U_{back}(j)\cL_{\e_x'} \left(\D t\parc{\e_x'(j)}{u_3}\right) U_{forw}(j)\r^{vec}_0.
\ee
In the case of limited pulse length, we have
\be
\parc{\r^{vec}_{N}}{u_{1,2}(s)}=\sum_{j=sn+1}^{(s+1)n} U_{back}(j)\cL_{\e_x'} \left(\D t\parc{\e_x'(j)}{u_{1,2}(s)}\right) U_{forw}(j)\r^{vec}_0,
\ee
where $n=l/\D t$, $l$ is the pulse length and $s=0, 1, \cdots ,T/l-1$.

The derivative of $\bar p_4$ with respect to the control parameter $u_k(j)$ is
\be
\parc{\bar p_4}{u_k(j)}=Tr[\hat{P}_{A_2}\parc{\r_{ave}}{u_k(j)}],
\ee
 where $\hat{P}_{A_2}=|A_2\rangle \langle A_2|$ and $\rho_{ave}=(\rho_N+\cdots+\rho_1)/N$.
Similarly we have
\be\label{derivativeave}
\parc{\r^{vec}_{ave}}{u_k(j)}=\frac{1}{N}\left(\sum_{i=j+1}^N \prod_{h=j+1}^i  e^{\cL_h\D t}+\mathbbm{1}\right)  \parc{e^{\cL_j \Delta t}}{u_k(j)} \r^{vec}_{j-1},
\ee
where we assign $\sum_{i=j+1}^N \prod_{h=j+1}^i  e^{\cL_h\D t}=0$ when $j=N$.
We define the following stairway operator as
\be
U_{stair}(j)=\sum_{i=j+1}^N \prod_{h=j+1}^i  e^{\cL_h\D t}+\mathbbm{1}
\ee
 and substitute Eq.(\ref{pluglast}) into the righthand side of Eq.(\ref{derivativeave}), which leads to
 \be
\parc{\r^{vec}_{ave}}{u_k(j)}=\frac{ \D t }{N}U_{stair}(j)\cL_{\e_x'} \left(\parc{\e_x'(j)}{u_k(j)}\right) U_{forw}(j)\r_0^{vec}.
\ee
For $u_3\equiv\D$,
\be
\parc{\r^{vec}_{ave}}{u_3}=\sum_j^N\frac{ \D t }{N}U_{stair}(j)\cL_{\e_x'} \left(\parc{\e_x'(j)}{u_3}\right) U_{forw}(j)\r_0^{vec}.
\ee
For the limited pulse length case
 \be
\parc{\r^{vec}_{ave}}{u_{1,2}(s)}=\sum_{j=sn+1}^{(s+1)n} \frac{ \D t }{N}U_{stair}(j)\cL_{\e_x'} \left(\parc{\e_x'(j)}{u_{1,2}(s)}\right) U_{forw}(j)\r_0^{vec}.
\ee
For each parameter specifically, we have
\be
\parc{\e_x'(j)}{\O_i(j)}=\cos\left[(\d_i+\Delta) t_j\right], i=1,2. ,
\ee
and
\be
\parc{\e_x'(j)}{\D}=-\Omega_1(j)\sin\left[(\d_1+\Delta) t_j\right] t_j-\Omega_2(j)\sin\left[(\d_2+\Delta) t_j\right] t_j.
\ee
Finally, the derivative of $E$ with respect to the control parameter $u_k(j)$ is simply given by
\be
\parc{E}{\O_i(j)}=2\O_i(j), i=1,2. ,
\ee
and
\be
\parc{E}{\D}=0.
\ee

\begin{figure}[b]
\hspace{0.1cm}
\includegraphics[width=1\columnwidth]{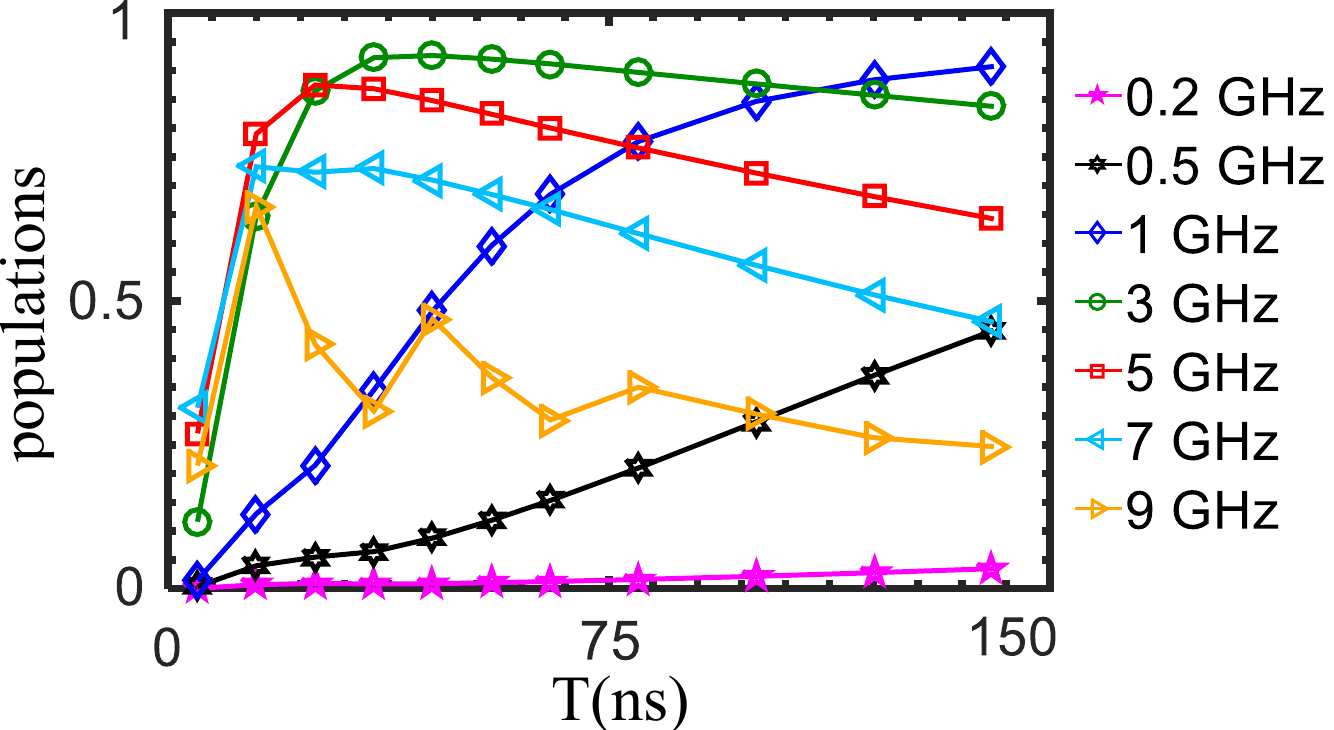}\\
 \caption{(Color online) The final population of the target state $\ket{+1}$ for the 10-level model with all dissipations as a function of the total evolution time $T$ of the STIRAP process. The peak amplitude of the control laser field is: $a=0.2$ GHz (pink star), $a=0.5$ GHz (black hexagonal), $a=1$ GHz (blue diamond), $3$ GHz (green circle), $5$ GHz (red square),  $7$ GHz (cyan left triangle),  $9$ GHz (gold right triangle). The other parameters are $\sigma=T/10$ and $\mu_{\pm}=T/2\pm\sigma$. The decay rates are presented in Table \ref{table}.
 }\label{figureapp}
\end{figure}

\begin{figure*}[t]
\hspace{-0.1cm}
\includegraphics[width=2\columnwidth]{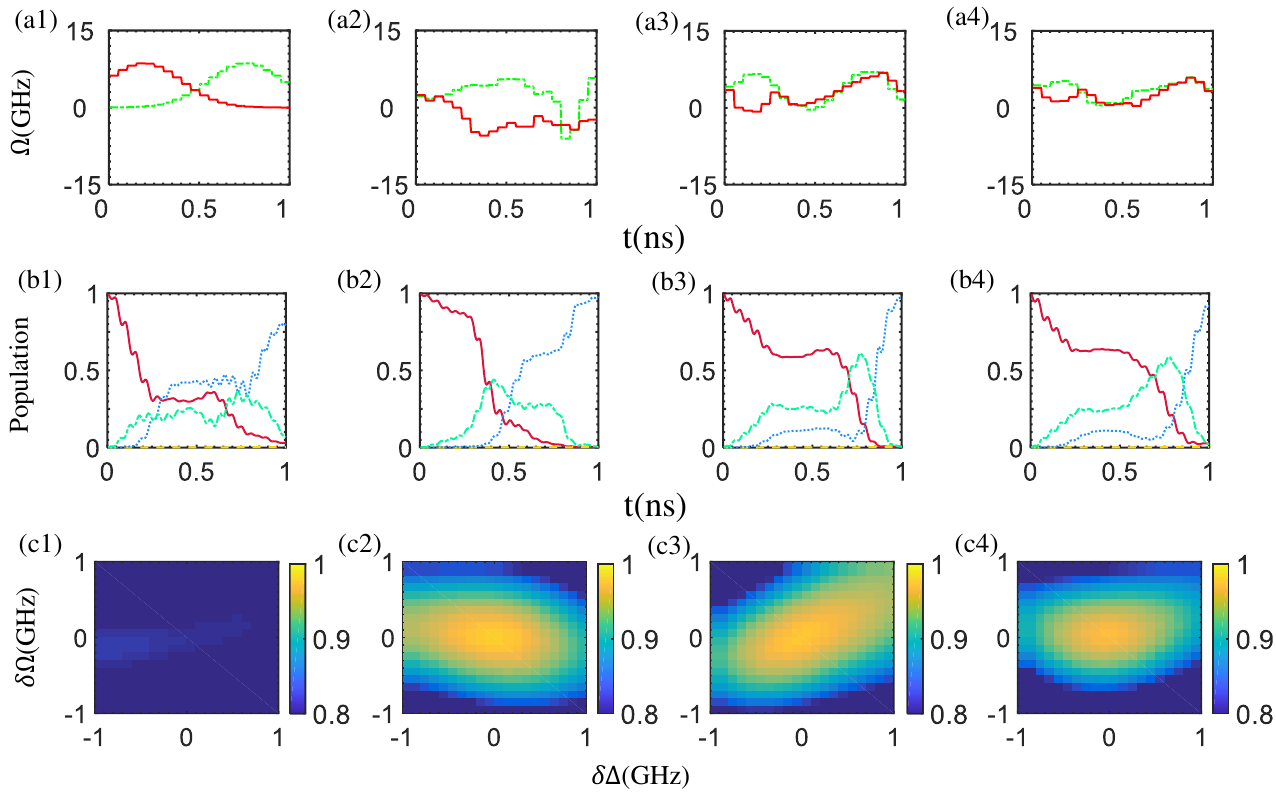}
\caption{(Color online) The details of the optimisation results with a time resolution of $0.05ns$. (a1-a4) show the amplitudes of the control laser field $\O_1(t)$ (green dashed line) and $\O_2(t)$ (red solid line) for the total evolution time $T=1$ ns (see Fig.\ref{fig4}) that are obtained by using the optimisation methods in Table~\ref{table1}.  (b1-b4) The dynamic evolution of the population of the state $\ket{-1}$ (red solid line), $\ket{0}$ (yellow dashed line), $\ket{+1}$ (blue dotted line) and $\ket{A_2}$ (green dashed-dotted line) for the corresponding four optimisation results with a total evolution time $T=1$ ns respectively. (c1-c4) The final population of the target state $\ket{+1}$ versus the amplitude systematic errors $\delta \O$ and the frequency detuning $\delta \Delta$ of the optimised control laser field. The four columns of figures correspond to the results using four optimisation methods: adiabatic Nelder-Mead (a1, b1, c1), adiabatic GRAPE (a2, b2, c2), Rabi resonant GRAPE (a3, b3, c3) and Rabi detuning GRAPE (a4, b4, c4), respectively. The optimal parameter of the detuning in the Rabi-detuning GRAPE method is $\D=-0.6519$ GHz.
}\label{figureapp2}
\end{figure*}

\section{Map to the Interaction picture}
In the laboratory frame the driven Hamiltonian is
\be
\begin{aligned}
H_c &= i\epsilon_x\ketbra{1}{4}-i\epsilon_x\ketbra{1}{5}-i\epsilon_x\ketbra{1}{8}\\
&-i\epsilon_x\ketbra{1}{9}+2\epsilon_x\ketbra{2}{7}-i\epsilon_x\ketbra{3}{4}\\
&-i\epsilon_x\ketbra{3}{5}+i\epsilon_x\ketbra{3}{8}+i\epsilon_x\ketbra{3}{9}+h.c.,\\
\end{aligned}
\ee
 with the driving field $\epsilon_x=\overline{\O}_1 \cos(\omega_1 t)+\overline{\O}_2 \cos(\omega_2 t)$, where $\overline{\O}_1$ and $\overline{\O}_2$ are the amplitudes with $\omega_1$ and $\omega_2$ the frequencies of the laser field. In the  resonant case $\omega_1$ ($\omega_2$) matches the energy gap between $\ket{A_2}$ and $\ket{-1}$ ($\ket{+1}$).
In the interaction picture with respect to Hamiltonian
\be
H_{\alpha}=H_0\equiv\sum_{i=1}^{10}E_i \ketbra{i}{i},
\ee
the driven Hamiltonian becomes
\be\label{eq:App_IntH}
\begin{aligned}
&e^{iH_{\alpha}t}H_c e^{-iH_{\alpha}t}=\\
 &ie^{i\D_{14}t}\epsilon_x\ketbra{1}{4}-ie^{i\D_{15}t}\epsilon_x\ketbra{1}{5}-ie^{i\D_{18}t}\epsilon_x\ketbra{1}{8}\\
&-ie^{i\D_{19}t}\epsilon_x\ketbra{1}{9}+2e^{i\D_{27}t}\epsilon_x\ketbra{2}{7}-ie^{i\D_{34}t}\epsilon_x\ketbra{3}{4}\\
&-ie^{i\D_{35}t}\epsilon_x\ketbra{3}{5}+ie^{i\D_{38}t}\epsilon_x\ketbra{3}{8}+ie^{i\D_{39}t}\epsilon_x\ketbra{3}{9}+h.c.,\\
\end{aligned}
\ee
where we denote $\D_{ij}=E_i-E_j$. For each matrix element above, the term $e^{\D_{ij}t}\epsilon_x$ is approximated to be
\be\label{eq:App_appr}
 e^{\D_{ij}t}\epsilon_x \approx \frac{\overline{\O}_1}{2}\cos\left[(\D_{ij}+\omega_1)t\right]+\frac{\overline{\O}_2}{2}\cos\left[(\D_{ij}+\omega_2)t\right],
 \ee
 where the terms $\cos\left[(\D_{ij}-\omega_1)t\right]$ and  $\cos\left[(\D_{ij}-\omega_2)t\right]$ are eliminated since $\omega_1=-\D_{14}$ and $\omega_2=-\D_{34}$.

To further simplify Eq.(\ref{eq:App_IntH}), instead of using $H_{\alpha}=H_0$, we choose $H_{\alpha}$ as
\be
H_{\alpha}=H_{E_g}\equiv E_g \sum_{i=4}^{9}\ketbra{i}{i},
\ee
such that the right-hand side of Eq.(\ref{eq:App_appr}) becomes the same as
\be
\frac{\overline{\O}_1}{2}\cos\left[(\omega_1-E_g)t\right]+\frac{\overline{\O}_2}{2}\cos\left[(\omega_2-E_g)t\right]
\ee
for all $i$ and $j$. The other terms in $H_0$ are left along with $V$ and eventually we get the Hamiltonian in the interaction picture as
\be
H_I=e^{iH_{E_g}t}(H_{gs}+H_{eg}+V-H_{E_g})e^{-iH_{E_g}t},
\ee
which is Eq.(\ref{eq:IntH}) in the main text.

\begin{figure}[t]
\hspace{-0.1cm}
\includegraphics[width=1\columnwidth]{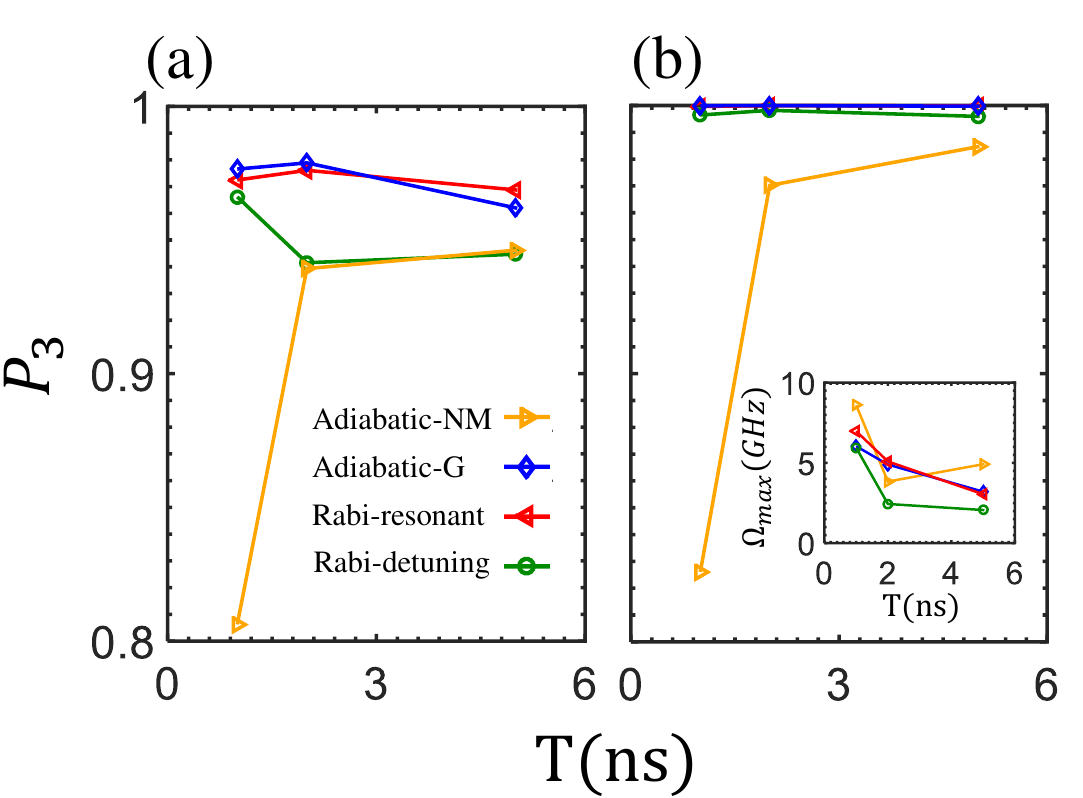}
\caption{(Color online) The optimal results with a time resolution of $0.05ns$ and different evolution times $T$ using four optimisation methods (as listed in Table~\ref{table1}) of the 10-level model of NV center spin. Each data is based on $500$ random initial points.  (a) shows the final population $p_3$ of the target state $\ket{+1}$ under the influence of dissipations. (b) shows the final population $p_3$ of the target state $\ket{+1}$ without dissipation. The inset shows the corresponding maximum laser field amplitudes that achieve the optimal fidelity of coherent state transfer. The results from four types optimal methods are presented in different symbols: (1) adiabatic Nelder-Mead (yellow triangle), (2) adiabatic GRAPE (blue diamond), (3) Rabi resonant GRAPE (red triangle) and (4) Rabi detuning GRAPE (green circle).
}\label{figureapp3}
\end{figure}

\section{More details on optimisation results }

\subsection{Effect of laser intensity on STIRAP}\label{Morecharacters}

To investigate the effect of the parameter of the laser intensity $a$, we calculate the performance of STIRAP process for different parameters $a$ with $a=0.2, 0.5, 1, 3, 5, 7, 9$ GHz, as shown in Fig. \ref{figureapp}. It shows that too weaker or stronger pulses will result in a worse performance.
\\

\subsection{Optimal results with a worse time resolution in modulation}\label{STIRAP1ns}
The laser pulse shape can be modulated by a $10$ Gs/s arbitrary wave generator (AWG) Waveform Conversion with a time resolution about $0.1$ ns \cite{Brian.NP2016}. An AWG with higher time resolution (for example, $24$ Gs/s \cite{Scheuer.njp2014}) will enable an even better modulation of the laser pulse (about $0.042$ ns). Here, we provide numerical simulation result which shows that the optimal control still works with a total evolution time $T=1$ ns with a time resolution in the modulation of $0.05$ ns, as shown in Fig. \ref{figureapp2}. The optimisation conditions are given in Table.\ref{table1}. The final populations of the target states $\ket {+1}$ are $0.8062$, $0.9765$, $0.9724$ and $0.9662$ from left to right, respectively. Fidelities and corresponding maximal field amplitudes of optimal results with evolution time of $1$ ns, $2$ ns and $5$ ns are showed in Fig. \ref{figureapp3}.

\bibliography{OptimalBib}

\begin{thebibliography}{57}%
\makeatletter
\providecommand \@ifxundefined [1]{%
 \@ifx{#1\undefined}
}%
\providecommand \@ifnum [1]{%
 \ifnum #1\expandafter \@firstoftwo
 \else \expandafter \@secondoftwo
 \fi
}%
\providecommand \@ifx [1]{%
 \ifx #1\expandafter \@firstoftwo
 \else \expandafter \@secondoftwo
 \fi
}%
\providecommand \natexlab [1]{#1}%
\providecommand \enquote  [1]{``#1''}%
\providecommand \bibnamefont  [1]{#1}%
\providecommand \bibfnamefont [1]{#1}%
\providecommand \citenamefont [1]{#1}%
\providecommand \href@noop [0]{\@secondoftwo}%
\providecommand \href [0]{\begingroup \@sanitize@url \@href}%
\providecommand \@href[1]{\@@startlink{#1}\@@href}%
\providecommand \@@href[1]{\endgroup#1\@@endlink}%
\providecommand \@sanitize@url [0]{\catcode `\\12\catcode `\$12\catcode
  `\&12\catcode `\#12\catcode `\^12\catcode `\_12\catcode `\%12\relax}%
\providecommand \@@startlink[1]{}%
\providecommand \@@endlink[0]{}%
\providecommand \url  [0]{\begingroup\@sanitize@url \@url }%
\providecommand \@url [1]{\endgroup\@href {#1}{\urlprefix }}%
\providecommand \urlprefix  [0]{URL }%
\providecommand \Eprint [0]{\href }%
\providecommand \doibase [0]{http://dx.doi.org/}%
\providecommand \selectlanguage [0]{\@gobble}%
\providecommand \bibinfo  [0]{\@secondoftwo}%
\providecommand \bibfield  [0]{\@secondoftwo}%
\providecommand \translation [1]{[#1]}%
\providecommand \BibitemOpen [0]{}%
\providecommand \bibitemStop [0]{}%
\providecommand \bibitemNoStop [0]{.\EOS\space}%
\providecommand \EOS [0]{\spacefactor3000\relax}%
\providecommand \BibitemShut  [1]{\csname bibitem#1\endcsname}%
\let\auto@bib@innerbib\@empty
\bibitem [{\citenamefont {Gruber}\ \emph {et~al.}(1997)\citenamefont {Gruber},
  \citenamefont {Dr{\"a}benstedt}, \citenamefont {Tietz}, \citenamefont
  {Fleury}, \citenamefont {Wrachtrup},\ and\ \citenamefont
  {Borczyskowski}}]{Gruber.S1997}%
  \BibitemOpen
  \bibfield  {author} {\bibinfo {author} {\bibfnamefont {A.}~\bibnamefont
  {Gruber}}, \bibinfo {author} {\bibfnamefont {A.}~\bibnamefont
  {Dr{\"a}benstedt}}, \bibinfo {author} {\bibfnamefont {C.}~\bibnamefont
  {Tietz}}, \bibinfo {author} {\bibfnamefont {L.}~\bibnamefont {Fleury}},
  \bibinfo {author} {\bibfnamefont {J.}~\bibnamefont {Wrachtrup}}, \ and\
  \bibinfo {author} {\bibfnamefont {C.~von}\ \bibnamefont {Borczyskowski}},\
  }\bibfield  {title} {\enquote {\bibinfo {title} {Scanning confocal optical
  microscopy and magnetic resonance on single defect centers},}\ }\href
  {\doibase 10.1126/science.276.5321.2012} {\bibfield  {journal} {\bibinfo
  {journal} {Science}\ }\textbf {\bibinfo {volume} {276}},\ \bibinfo {pages}
  {2012--2014} (\bibinfo {year} {1997})}\BibitemShut {NoStop}%
\bibitem [{\citenamefont {Jelezko}\ \emph
  {et~al.}(2004{\natexlab{a}})\citenamefont {Jelezko}, \citenamefont {Gaebel},
  \citenamefont {Popa}, \citenamefont {Gruber},\ and\ \citenamefont
  {Wrachtrup}}]{Jelezko.PRL2004}%
  \BibitemOpen
  \bibfield  {author} {\bibinfo {author} {\bibfnamefont {F.}~\bibnamefont
  {Jelezko}}, \bibinfo {author} {\bibfnamefont {T.}~\bibnamefont {Gaebel}},
  \bibinfo {author} {\bibfnamefont {I.}~\bibnamefont {Popa}}, \bibinfo {author}
  {\bibfnamefont {A.}~\bibnamefont {Gruber}}, \ and\ \bibinfo {author}
  {\bibfnamefont {J.}~\bibnamefont {Wrachtrup}},\ }\bibfield  {title} {\enquote
  {\bibinfo {title} {Observation of coherent oscillations in a single electron
  spin},}\ }\href {\doibase 10.1103/PhysRevLett.92.076401} {\bibfield
  {journal} {\bibinfo  {journal} {Phys. Rev. Lett.}\ }\textbf {\bibinfo
  {volume} {92}},\ \bibinfo {pages} {076401} (\bibinfo {year}
  {2004}{\natexlab{a}})}\BibitemShut {NoStop}%
\bibitem [{\citenamefont {Jelezko}\ and\ \citenamefont
  {Wrachtrup}(2006)}]{Jelezko.PSSA2006}%
  \BibitemOpen
  \bibfield  {author} {\bibinfo {author} {\bibfnamefont {F.}~\bibnamefont
  {Jelezko}}\ and\ \bibinfo {author} {\bibfnamefont {J.}~\bibnamefont
  {Wrachtrup}},\ }\bibfield  {title} {\enquote {\bibinfo {title} {Single defect
  centres in diamond: A review},}\ }\href {\doibase 10.1002/pssa.200671403}
  {\bibfield  {journal} {\bibinfo  {journal} {Physica Status Solidi A}\
  }\textbf {\bibinfo {volume} {203}},\ \bibinfo {pages} {3207--3225} (\bibinfo
  {year} {2006})}\BibitemShut {NoStop}%
\bibitem [{\citenamefont {Doherty}\ \emph {et~al.}(2013)\citenamefont
  {Doherty}, \citenamefont {Manson}, \citenamefont {Delaney}, \citenamefont
  {Jelezko}, \citenamefont {Wrachtrup},\ and\ \citenamefont
  {Hollenberg}}]{Dohertya.PR2013}%
  \BibitemOpen
  \bibfield  {author} {\bibinfo {author} {\bibfnamefont {M.~W.}\ \bibnamefont
  {Doherty}}, \bibinfo {author} {\bibfnamefont {N.~B.}\ \bibnamefont {Manson}},
  \bibinfo {author} {\bibfnamefont {P.}~\bibnamefont {Delaney}}, \bibinfo
  {author} {\bibfnamefont {F.}~\bibnamefont {Jelezko}}, \bibinfo {author}
  {\bibfnamefont {J.}~\bibnamefont {Wrachtrup}}, \ and\ \bibinfo {author}
  {\bibfnamefont {L.~C.L.}\ \bibnamefont {Hollenberg}},\ }\bibfield  {title}
  {\enquote {\bibinfo {title} {The nitrogen-vacancy colour centre in
  diamond},}\ }\href {\doibase https://doi.org/10.1016/j.physrep.2013.02.001}
  {\bibfield  {journal} {\bibinfo  {journal} {Physics Reports}\ }\textbf
  {\bibinfo {volume} {528}},\ \bibinfo {pages} {1 -- 45} (\bibinfo {year}
  {2013})}\BibitemShut {NoStop}%
\bibitem [{\citenamefont {Balasubramanian}\ \emph {et~al.}(2009)\citenamefont
  {Balasubramanian}, \citenamefont {Neumann}, \citenamefont {Twitchen},
  \citenamefont {Markham}, \citenamefont {Kolesov}, \citenamefont {Mizuochi},
  \citenamefont {Isoya}, \citenamefont {Achard}, \citenamefont {Beck},
  \citenamefont {Tissler}, \citenamefont {Jacques}, \citenamefont {Hemmer},
  \citenamefont {Jelezko},\ and\ \citenamefont
  {Wrachtrup}}]{Balasubramanian.NM2009}%
  \BibitemOpen
  \bibfield  {author} {\bibinfo {author} {\bibfnamefont {G.}~\bibnamefont
  {Balasubramanian}}, \bibinfo {author} {\bibfnamefont {P.}~\bibnamefont
  {Neumann}}, \bibinfo {author} {\bibfnamefont {D.}~\bibnamefont {Twitchen}},
  \bibinfo {author} {\bibfnamefont {M.}~\bibnamefont {Markham}}, \bibinfo
  {author} {\bibfnamefont {R.}~\bibnamefont {Kolesov}}, \bibinfo {author}
  {\bibfnamefont {N.}~\bibnamefont {Mizuochi}}, \bibinfo {author}
  {\bibfnamefont {J.}~\bibnamefont {Isoya}}, \bibinfo {author} {\bibfnamefont
  {J.}~\bibnamefont {Achard}}, \bibinfo {author} {\bibfnamefont
  {J.}~\bibnamefont {Beck}}, \bibinfo {author} {\bibfnamefont {J.}~\bibnamefont
  {Tissler}}, \bibinfo {author} {\bibfnamefont {V.}~\bibnamefont {Jacques}},
  \bibinfo {author} {\bibfnamefont {Philip~R.}\ \bibnamefont {Hemmer}},
  \bibinfo {author} {\bibfnamefont {F.}~\bibnamefont {Jelezko}}, \ and\
  \bibinfo {author} {\bibfnamefont {J.}~\bibnamefont {Wrachtrup}},\ }\bibfield
  {title} {\enquote {\bibinfo {title} {Ultralong spin coherence time in
  isotopically engineered diamond},}\ }\href {https://doi.org/10.1038/nmat2420}
  {\bibfield  {journal} {\bibinfo  {journal} {Nature Materials}\ }\textbf
  {\bibinfo {volume} {8}},\ \bibinfo {pages} {383} (\bibinfo {year}
  {2009})}\BibitemShut {NoStop}%
\bibitem [{\citenamefont {Maurer}\ \emph {et~al.}(2012)\citenamefont {Maurer},
  \citenamefont {Kucsko}, \citenamefont {Latta}, \citenamefont {Jiang},
  \citenamefont {Yao}, \citenamefont {Bennett}, \citenamefont {Pastawski},
  \citenamefont {Hunger}, \citenamefont {Chisholm}, \citenamefont {Markham},
  \citenamefont {Twitchen}, \citenamefont {Cirac},\ and\ \citenamefont
  {Lukin}}]{Maurer.S2012}%
  \BibitemOpen
  \bibfield  {author} {\bibinfo {author} {\bibfnamefont {P.~C.}\ \bibnamefont
  {Maurer}}, \bibinfo {author} {\bibfnamefont {G.}~\bibnamefont {Kucsko}},
  \bibinfo {author} {\bibfnamefont {C.}~\bibnamefont {Latta}}, \bibinfo
  {author} {\bibfnamefont {L.}~\bibnamefont {Jiang}}, \bibinfo {author}
  {\bibfnamefont {N.~Y.}\ \bibnamefont {Yao}}, \bibinfo {author} {\bibfnamefont
  {S.~D.}\ \bibnamefont {Bennett}}, \bibinfo {author} {\bibfnamefont
  {F.}~\bibnamefont {Pastawski}}, \bibinfo {author} {\bibfnamefont
  {D.}~\bibnamefont {Hunger}}, \bibinfo {author} {\bibfnamefont
  {N.}~\bibnamefont {Chisholm}}, \bibinfo {author} {\bibfnamefont
  {M.}~\bibnamefont {Markham}}, \bibinfo {author} {\bibfnamefont {D.~J.}\
  \bibnamefont {Twitchen}}, \bibinfo {author} {\bibfnamefont {J.~I.}\
  \bibnamefont {Cirac}}, \ and\ \bibinfo {author} {\bibfnamefont {M.~D.}\
  \bibnamefont {Lukin}},\ }\bibfield  {title} {\enquote {\bibinfo {title}
  {Room-temperature quantum bit memory exceeding one second},}\ }\href
  {\doibase 10.1126/science.1220513} {\bibfield  {journal} {\bibinfo  {journal}
  {Science}\ }\textbf {\bibinfo {volume} {336}},\ \bibinfo {pages} {1283--1286}
  (\bibinfo {year} {2012})}\BibitemShut {NoStop}%
\bibitem [{\citenamefont {Wrachtrup}\ and\ \citenamefont
  {Jelezko}(2006)}]{Wrachtrup.J2006}%
  \BibitemOpen
  \bibfield  {author} {\bibinfo {author} {\bibfnamefont {J.}~\bibnamefont
  {Wrachtrup}}\ and\ \bibinfo {author} {\bibfnamefont {F.}~\bibnamefont
  {Jelezko}},\ }\bibfield  {title} {\enquote {\bibinfo {title} {Processing
  quantum information in diamond},}\ }\href {\doibase
  10.1088/0953-8984/18/21/s08} {\bibfield  {journal} {\bibinfo  {journal}
  {Journal of Physics: Condensed Matter}\ }\textbf {\bibinfo {volume} {18}},\
  \bibinfo {pages} {S807--S824} (\bibinfo {year} {2006})}\BibitemShut {NoStop}%
\bibitem [{\citenamefont {van~der Sar}\ \emph {et~al.}(2012)\citenamefont
  {van~der Sar}, \citenamefont {Wang}, \citenamefont {Blok}, \citenamefont
  {Bernien}, \citenamefont {Taminiau}, \citenamefont {Toyli}, \citenamefont
  {Lidar}, \citenamefont {Awschalom}, \citenamefont {Hanson},\ and\
  \citenamefont {Dobrovitski}}]{Sar.N2012}%
  \BibitemOpen
  \bibfield  {author} {\bibinfo {author} {\bibfnamefont {T.}~\bibnamefont
  {van~der Sar}}, \bibinfo {author} {\bibfnamefont {Z.~H.}\ \bibnamefont
  {Wang}}, \bibinfo {author} {\bibfnamefont {M.~S.}\ \bibnamefont {Blok}},
  \bibinfo {author} {\bibfnamefont {H.}~\bibnamefont {Bernien}}, \bibinfo
  {author} {\bibfnamefont {T.~H.}\ \bibnamefont {Taminiau}}, \bibinfo {author}
  {\bibfnamefont {D.~M.}\ \bibnamefont {Toyli}}, \bibinfo {author}
  {\bibfnamefont {D.~A.}\ \bibnamefont {Lidar}}, \bibinfo {author}
  {\bibfnamefont {D.~D.}\ \bibnamefont {Awschalom}}, \bibinfo {author}
  {\bibfnamefont {R.}~\bibnamefont {Hanson}}, \ and\ \bibinfo {author}
  {\bibfnamefont {V.~V.}\ \bibnamefont {Dobrovitski}},\ }\bibfield  {title}
  {\enquote {\bibinfo {title} {Decoherence-protected quantum gates for a hybrid
  solid-state spin register},}\ }\href {https://doi.org/10.1038/nature10900}
  {\bibfield  {journal} {\bibinfo  {journal} {Nature}\ }\textbf {\bibinfo
  {volume} {484}},\ \bibinfo {pages} {82} (\bibinfo {year} {2012})}\BibitemShut
  {NoStop}%
\bibitem [{\citenamefont {Shi}\ \emph {et~al.}(2010)\citenamefont {Shi},
  \citenamefont {Rong}, \citenamefont {Xu}, \citenamefont {Wang}, \citenamefont
  {Wu}, \citenamefont {Chong}, \citenamefont {Peng}, \citenamefont {Kniepert},
  \citenamefont {Schoenfeld}, \citenamefont {Harneit}, \citenamefont {Feng},\
  and\ \citenamefont {Du}}]{Shi.PRL2010}%
  \BibitemOpen
  \bibfield  {author} {\bibinfo {author} {\bibfnamefont {F.}~\bibnamefont
  {Shi}}, \bibinfo {author} {\bibfnamefont {X.}~\bibnamefont {Rong}}, \bibinfo
  {author} {\bibfnamefont {N.}~\bibnamefont {Xu}}, \bibinfo {author}
  {\bibfnamefont {Y.}~\bibnamefont {Wang}}, \bibinfo {author} {\bibfnamefont
  {J.}~\bibnamefont {Wu}}, \bibinfo {author} {\bibfnamefont {B.}~\bibnamefont
  {Chong}}, \bibinfo {author} {\bibfnamefont {X.}~\bibnamefont {Peng}},
  \bibinfo {author} {\bibfnamefont {J.}~\bibnamefont {Kniepert}}, \bibinfo
  {author} {\bibfnamefont {R.-S.}\ \bibnamefont {Schoenfeld}}, \bibinfo
  {author} {\bibfnamefont {W.}~\bibnamefont {Harneit}}, \bibinfo {author}
  {\bibfnamefont {M.}~\bibnamefont {Feng}}, \ and\ \bibinfo {author}
  {\bibfnamefont {J.}~\bibnamefont {Du}},\ }\bibfield  {title} {\enquote
  {\bibinfo {title} {Room-temperature implementation of the deutsch-jozsa
  algorithm with a single electronic spin in diamond},}\ }\href {\doibase
  10.1103/PhysRevLett.105.040504} {\bibfield  {journal} {\bibinfo  {journal}
  {Phys. Rev. Lett.}\ }\textbf {\bibinfo {volume} {105}},\ \bibinfo {pages}
  {040504} (\bibinfo {year} {2010})}\BibitemShut {NoStop}%
\bibitem [{\citenamefont {Cai}\ \emph {et~al.}(2013)\citenamefont {Cai},
  \citenamefont {Retzker}, \citenamefont {Jelezko},\ and\ \citenamefont
  {Plenio}}]{Cai2013}%
  \BibitemOpen
  \bibfield  {author} {\bibinfo {author} {\bibfnamefont {J.}~\bibnamefont
  {Cai}}, \bibinfo {author} {\bibfnamefont {A.}~\bibnamefont {Retzker}},
  \bibinfo {author} {\bibfnamefont {F.}~\bibnamefont {Jelezko}}, \ and\
  \bibinfo {author} {\bibfnamefont {M.~B.}\ \bibnamefont {Plenio}},\ }\bibfield
   {title} {\enquote {\bibinfo {title} {A large-scale quantum simulator on a
  diamond surface at room temperature},}\ }\href {\doibase 10.1038/nphys2519}
  {\bibfield  {journal} {\bibinfo  {journal} {Nature Physics}\ }\textbf
  {\bibinfo {volume} {9}},\ \bibinfo {pages} {168} (\bibinfo {year}
  {2013})}\BibitemShut {NoStop}%
\bibitem [{\citenamefont {Arroyo-Camejo}\ \emph {et~al.}(2014)\citenamefont
  {Arroyo-Camejo}, \citenamefont {Lazariev}, \citenamefont {Hell},\ and\
  \citenamefont {Balasubramanian}}]{Arroyo.NC2014}%
  \BibitemOpen
  \bibfield  {author} {\bibinfo {author} {\bibfnamefont {S.}~\bibnamefont
  {Arroyo-Camejo}}, \bibinfo {author} {\bibfnamefont {A.}~\bibnamefont
  {Lazariev}}, \bibinfo {author} {\bibfnamefont {Stefan~W.}\ \bibnamefont
  {Hell}}, \ and\ \bibinfo {author} {\bibfnamefont {G.}~\bibnamefont
  {Balasubramanian}},\ }\bibfield  {title} {\enquote {\bibinfo {title} {Room
  temperature high-fidelity holonomic single-qubit gate on a solid-state
  spin},}\ }\href {https://doi.org/10.1038/ncomms5870} {\bibfield  {journal}
  {\bibinfo  {journal} {Nature Communications}\ }\textbf {\bibinfo {volume}
  {5}},\ \bibinfo {pages} {4870} (\bibinfo {year} {2014})}\BibitemShut
  {NoStop}%
\bibitem [{\citenamefont {Barfuss}\ \emph {et~al.}(2015)\citenamefont
  {Barfuss}, \citenamefont {Teissier}, \citenamefont {Neu}, \citenamefont
  {Nunnenkamp},\ and\ \citenamefont {Maletinsky}}]{Barfuss2015}%
  \BibitemOpen
  \bibfield  {author} {\bibinfo {author} {\bibfnamefont {A.}~\bibnamefont
  {Barfuss}}, \bibinfo {author} {\bibfnamefont {J.}~\bibnamefont {Teissier}},
  \bibinfo {author} {\bibfnamefont {E.}~\bibnamefont {Neu}}, \bibinfo {author}
  {\bibfnamefont {A.}~\bibnamefont {Nunnenkamp}}, \ and\ \bibinfo {author}
  {\bibfnamefont {P.}~\bibnamefont {Maletinsky}},\ }\bibfield  {title}
  {\enquote {\bibinfo {title} {Strong mechanical driving of a single electron
  spin},}\ }\href {\doibase https://doi.org/10.1038/nphys3411} {\bibfield
  {journal} {\bibinfo  {journal} {Nature physics}\ }\textbf {\bibinfo {volume}
  {11}},\ \bibinfo {pages} {820--824} (\bibinfo {year} {2015})}\BibitemShut
  {NoStop}%
\bibitem [{\citenamefont {Shu}\ \emph {et~al.}(2018)\citenamefont {Shu},
  \citenamefont {Liu}, \citenamefont {Cao}, \citenamefont {Yang}, \citenamefont
  {Zhang}, \citenamefont {Plenio}, \citenamefont {Jelezko},\ and\ \citenamefont
  {Cai}}]{Shu.prl2018}%
  \BibitemOpen
  \bibfield  {author} {\bibinfo {author} {\bibfnamefont {Z.}~\bibnamefont
  {Shu}}, \bibinfo {author} {\bibfnamefont {Y.}~\bibnamefont {Liu}}, \bibinfo
  {author} {\bibfnamefont {Q.}~\bibnamefont {Cao}}, \bibinfo {author}
  {\bibfnamefont {P.}~\bibnamefont {Yang}}, \bibinfo {author} {\bibfnamefont
  {S.}~\bibnamefont {Zhang}}, \bibinfo {author} {\bibfnamefont {Martin~B.}\
  \bibnamefont {Plenio}}, \bibinfo {author} {\bibfnamefont {F.}~\bibnamefont
  {Jelezko}}, \ and\ \bibinfo {author} {\bibfnamefont {J.}~\bibnamefont
  {Cai}},\ }\bibfield  {title} {\enquote {\bibinfo {title} {Observation of
  floquet raman transition in a driven solid-state spin system},}\ }\href
  {\doibase 10.1103/PhysRevLett.121.210501} {\bibfield  {journal} {\bibinfo
  {journal} {Phys. Rev. Lett.}\ }\textbf {\bibinfo {volume} {121}},\ \bibinfo
  {pages} {210501} (\bibinfo {year} {2018})}\BibitemShut {NoStop}%
\bibitem [{\citenamefont {Yu}\ \emph {et~al.}(2018)\citenamefont {Yu},
  \citenamefont {Yang}, \citenamefont {Gong}, \citenamefont {Cao},
  \citenamefont {Lu}, \citenamefont {Liu}, \citenamefont {Plenio},
  \citenamefont {Jelezko}, \citenamefont {Ozawa}, \citenamefont {Goldman},
  \citenamefont {Zhang},\ and\ \citenamefont {Cai}}]{Yu2018}%
  \BibitemOpen
  \bibfield  {author} {\bibinfo {author} {\bibfnamefont {Min}\ \bibnamefont
  {Yu}}, \bibinfo {author} {\bibfnamefont {Pengcheng}\ \bibnamefont {Yang}},
  \bibinfo {author} {\bibfnamefont {Musang}\ \bibnamefont {Gong}}, \bibinfo
  {author} {\bibfnamefont {Qingyun}\ \bibnamefont {Cao}}, \bibinfo {author}
  {\bibfnamefont {Qiuyu}\ \bibnamefont {Lu}}, \bibinfo {author} {\bibfnamefont
  {Haibin}\ \bibnamefont {Liu}}, \bibinfo {author} {\bibfnamefont {Martin~B.}\
  \bibnamefont {Plenio}}, \bibinfo {author} {\bibfnamefont {Fedor}\
  \bibnamefont {Jelezko}}, \bibinfo {author} {\bibfnamefont {Tomoki}\
  \bibnamefont {Ozawa}}, \bibinfo {author} {\bibfnamefont {Nathan}\
  \bibnamefont {Goldman}}, \bibinfo {author} {\bibfnamefont {Shaoliang}\
  \bibnamefont {Zhang}}, \ and\ \bibinfo {author} {\bibfnamefont {Jianming}\
  \bibnamefont {Cai}},\ }\bibfield  {title} {\enquote {\bibinfo {title}
  {Experimental measurement of the complete quantum geometry of a solid-state
  spin system},}\ }\href {https://arxiv.org/abs/1811.12840} {\bibfield
  {journal} {\bibinfo  {journal} {arXiv:}\ }\textbf {\bibinfo {volume}
  {1811}},\ \bibinfo {pages} {12840} (\bibinfo {year} {2018})}\BibitemShut
  {NoStop}%
\bibitem [{\citenamefont {Balasubramanian}\ \emph {et~al.}(2008)\citenamefont
  {Balasubramanian}, \citenamefont {Chan}, \citenamefont {Kolesov},
  \citenamefont {Al-Hmoud}, \citenamefont {Tisler}, \citenamefont {Shin},
  \citenamefont {Kim}, \citenamefont {Wojcik}, \citenamefont {Hemmer},
  \citenamefont {Krueger} \emph {et~al.}}]{Balasubramanian.N2008}%
  \BibitemOpen
  \bibfield  {author} {\bibinfo {author} {\bibfnamefont {G.}~\bibnamefont
  {Balasubramanian}}, \bibinfo {author} {\bibfnamefont {I.~Y.}\ \bibnamefont
  {Chan}}, \bibinfo {author} {\bibfnamefont {R.}~\bibnamefont {Kolesov}},
  \bibinfo {author} {\bibfnamefont {M.}~\bibnamefont {Al-Hmoud}}, \bibinfo
  {author} {\bibfnamefont {J.}~\bibnamefont {Tisler}}, \bibinfo {author}
  {\bibfnamefont {C.}~\bibnamefont {Shin}}, \bibinfo {author} {\bibfnamefont
  {C.}~\bibnamefont {Kim}}, \bibinfo {author} {\bibfnamefont {A.}~\bibnamefont
  {Wojcik}}, \bibinfo {author} {\bibfnamefont {P.~R.}\ \bibnamefont {Hemmer}},
  \bibinfo {author} {\bibfnamefont {A.}~\bibnamefont {Krueger}},  \emph
  {et~al.},\ }\bibfield  {title} {\enquote {\bibinfo {title} {Nanoscale imaging
  magnetometry with diamond spins under ambient conditions},}\ }\href
  {https://doi.org/10.1038/nature07278} {\bibfield  {journal} {\bibinfo
  {journal} {Nature}\ }\textbf {\bibinfo {volume} {455}},\ \bibinfo {pages}
  {648} (\bibinfo {year} {2008})}\BibitemShut {NoStop}%
\bibitem [{\citenamefont {Maze}\ \emph {et~al.}(2008)\citenamefont {Maze},
  \citenamefont {Stanwix}, \citenamefont {Hodges}, \citenamefont {Hong},
  \citenamefont {Taylor}, \citenamefont {Cappellaro}, \citenamefont {Jiang},
  \citenamefont {Dutt}, \citenamefont {Togan}, \citenamefont {Zibrov},
  \citenamefont {Yacoby}, \citenamefont {Walsworth},\ and\ \citenamefont
  {Lukin}}]{Maze.N2008}%
  \BibitemOpen
  \bibfield  {author} {\bibinfo {author} {\bibfnamefont {J.~R.}\ \bibnamefont
  {Maze}}, \bibinfo {author} {\bibfnamefont {P.~L.}\ \bibnamefont {Stanwix}},
  \bibinfo {author} {\bibfnamefont {J.~S.}\ \bibnamefont {Hodges}}, \bibinfo
  {author} {\bibfnamefont {S.}~\bibnamefont {Hong}}, \bibinfo {author}
  {\bibfnamefont {J.~M.}\ \bibnamefont {Taylor}}, \bibinfo {author}
  {\bibfnamefont {P.}~\bibnamefont {Cappellaro}}, \bibinfo {author}
  {\bibfnamefont {L.}~\bibnamefont {Jiang}}, \bibinfo {author} {\bibfnamefont
  {M.~V.~Gurudev}\ \bibnamefont {Dutt}}, \bibinfo {author} {\bibfnamefont
  {E.}~\bibnamefont {Togan}}, \bibinfo {author} {\bibfnamefont {A.~S.}\
  \bibnamefont {Zibrov}}, \bibinfo {author} {\bibfnamefont {A.}~\bibnamefont
  {Yacoby}}, \bibinfo {author} {\bibfnamefont {R.~L.}\ \bibnamefont
  {Walsworth}}, \ and\ \bibinfo {author} {\bibfnamefont {M.~D.}\ \bibnamefont
  {Lukin}},\ }\bibfield  {title} {\enquote {\bibinfo {title} {Nanoscale
  magnetic sensing with an individual electronic spin in diamond},}\ }\href
  {https://doi.org/10.1038/nature07279} {\bibfield  {journal} {\bibinfo
  {journal} {Nature}\ }\textbf {\bibinfo {volume} {455}},\ \bibinfo {pages}
  {644} (\bibinfo {year} {2008})}\BibitemShut {NoStop}%
\bibitem [{\citenamefont {Grinolds}\ \emph {et~al.}(2013)\citenamefont
  {Grinolds}, \citenamefont {Hong}, \citenamefont {Maletinsky}, \citenamefont
  {Luan}, \citenamefont {Lukin}, \citenamefont {Walsworth},\ and\ \citenamefont
  {Yacoby}}]{Grinolds.NP2013}%
  \BibitemOpen
  \bibfield  {author} {\bibinfo {author} {\bibfnamefont {M.~S.}\ \bibnamefont
  {Grinolds}}, \bibinfo {author} {\bibfnamefont {S.}~\bibnamefont {Hong}},
  \bibinfo {author} {\bibfnamefont {P.}~\bibnamefont {Maletinsky}}, \bibinfo
  {author} {\bibfnamefont {L.}~\bibnamefont {Luan}}, \bibinfo {author}
  {\bibfnamefont {M.~D.}\ \bibnamefont {Lukin}}, \bibinfo {author}
  {\bibfnamefont {R.~L.}\ \bibnamefont {Walsworth}}, \ and\ \bibinfo {author}
  {\bibfnamefont {A.}~\bibnamefont {Yacoby}},\ }\bibfield  {title} {\enquote
  {\bibinfo {title} {Nanoscale magnetic imaging of a single electron spin under
  ambient conditions},}\ }\href {https://doi.org/10.1038/nphys2543} {\bibfield
  {journal} {\bibinfo  {journal} {Nature Physics}\ }\textbf {\bibinfo {volume}
  {9}},\ \bibinfo {pages} {215} (\bibinfo {year} {2013})}\BibitemShut {NoStop}%
\bibitem [{\citenamefont {Cai}\ \emph {et~al.}(2014)\citenamefont {Cai},
  \citenamefont {Jelezko},\ and\ \citenamefont {Plenio}}]{Cai2014}%
  \BibitemOpen
  \bibfield  {author} {\bibinfo {author} {\bibfnamefont {J.}~\bibnamefont
  {Cai}}, \bibinfo {author} {\bibfnamefont {F.}~\bibnamefont {Jelezko}}, \ and\
  \bibinfo {author} {\bibfnamefont {Martin~B.}\ \bibnamefont {Plenio}},\
  }\bibfield  {title} {\enquote {\bibinfo {title} {Hybrid sensors based on
  colour centres in diamond and piezoactive layers},}\ }\href
  {https://doi.org/10.1038/ncomms5065} {\bibfield  {journal} {\bibinfo
  {journal} {Nature Communications}\ }\textbf {\bibinfo {volume} {5}},\
  \bibinfo {pages} {4065} (\bibinfo {year} {2014})}\BibitemShut {NoStop}%
\bibitem [{\citenamefont {M{\"u}ller}\ \emph {et~al.}(2014)\citenamefont
  {M{\"u}ller}, \citenamefont {Kong}, \citenamefont {Cai}, \citenamefont
  {Melentijevic}, \citenamefont {Stacey}, \citenamefont {Markham},
  \citenamefont {Twitchen}, \citenamefont {Isoya}, \citenamefont {Pezzagna},
  \citenamefont {Meijer}, \citenamefont {Du}, \citenamefont {Plenio},
  \citenamefont {Naydenov}, \citenamefont {McGuinness},\ and\ \citenamefont
  {Jelezko}}]{Muller.NC2014}%
  \BibitemOpen
  \bibfield  {author} {\bibinfo {author} {\bibfnamefont {C.}~\bibnamefont
  {M{\"u}ller}}, \bibinfo {author} {\bibfnamefont {X.}~\bibnamefont {Kong}},
  \bibinfo {author} {\bibfnamefont {J.-M.}\ \bibnamefont {Cai}}, \bibinfo
  {author} {\bibfnamefont {K.}~\bibnamefont {Melentijevic}}, \bibinfo {author}
  {\bibfnamefont {A.}~\bibnamefont {Stacey}}, \bibinfo {author} {\bibfnamefont
  {M.}~\bibnamefont {Markham}}, \bibinfo {author} {\bibfnamefont
  {D.}~\bibnamefont {Twitchen}}, \bibinfo {author} {\bibfnamefont
  {J.}~\bibnamefont {Isoya}}, \bibinfo {author} {\bibfnamefont
  {S.}~\bibnamefont {Pezzagna}}, \bibinfo {author} {\bibfnamefont
  {J.}~\bibnamefont {Meijer}}, \bibinfo {author} {\bibfnamefont
  {J.}~\bibnamefont {Du}}, \bibinfo {author} {\bibfnamefont {M.~B.}\
  \bibnamefont {Plenio}}, \bibinfo {author} {\bibfnamefont {B.}~\bibnamefont
  {Naydenov}}, \bibinfo {author} {\bibfnamefont {L.~P.}\ \bibnamefont
  {McGuinness}}, \ and\ \bibinfo {author} {\bibfnamefont {F.}~\bibnamefont
  {Jelezko}},\ }\bibfield  {title} {\enquote {\bibinfo {title} {Nuclear
  magnetic resonance spectroscopy with single spin sensitivity},}\ }\href
  {https://doi.org/10.1038/ncomms5703} {\bibfield  {journal} {\bibinfo
  {journal} {Nature Communications}\ }\textbf {\bibinfo {volume} {5}},\
  \bibinfo {pages} {4703} (\bibinfo {year} {2014})}\BibitemShut {NoStop}%
\bibitem [{\citenamefont {Sushkov}\ \emph {et~al.}(2014)\citenamefont
  {Sushkov}, \citenamefont {Chisholm}, \citenamefont {Lovchinsky},
  \citenamefont {Kubo}, \citenamefont {Lo}, \citenamefont {Bennett},
  \citenamefont {Hunger}, \citenamefont {Akimov}, \citenamefont {Walsworth},
  \citenamefont {Park},\ and\ \citenamefont {Lukin}}]{Sushkov2014}%
  \BibitemOpen
  \bibfield  {author} {\bibinfo {author} {\bibfnamefont {A.~O.}\ \bibnamefont
  {Sushkov}}, \bibinfo {author} {\bibfnamefont {N.}~\bibnamefont {Chisholm}},
  \bibinfo {author} {\bibfnamefont {I.}~\bibnamefont {Lovchinsky}}, \bibinfo
  {author} {\bibfnamefont {M.}~\bibnamefont {Kubo}}, \bibinfo {author}
  {\bibfnamefont {P.~K.}\ \bibnamefont {Lo}}, \bibinfo {author} {\bibfnamefont
  {S.~D.}\ \bibnamefont {Bennett}}, \bibinfo {author} {\bibfnamefont
  {D.}~\bibnamefont {Hunger}}, \bibinfo {author} {\bibfnamefont
  {A.}~\bibnamefont {Akimov}}, \bibinfo {author} {\bibfnamefont {R.~L.}\
  \bibnamefont {Walsworth}}, \bibinfo {author} {\bibfnamefont {H.}~\bibnamefont
  {Park}}, \ and\ \bibinfo {author} {\bibfnamefont {M.~D.}\ \bibnamefont
  {Lukin}},\ }\bibfield  {title} {\enquote {\bibinfo {title} {All-optical
  sensing of a single-molecule electron spin},}\ }\href {\doibase
  10.1021/nl502988n} {\bibfield  {journal} {\bibinfo  {journal} {Nano Letters}\
  }\textbf {\bibinfo {volume} {14}},\ \bibinfo {pages} {6443--6448} (\bibinfo
  {year} {2014})}\BibitemShut {NoStop}%
\bibitem [{\citenamefont {Shi}\ \emph {et~al.}(2015)\citenamefont {Shi},
  \citenamefont {Zhang}, \citenamefont {Wang}, \citenamefont {Sun},
  \citenamefont {Wang}, \citenamefont {Rong}, \citenamefont {Chen},
  \citenamefont {Ju}, \citenamefont {Reinhard}, \citenamefont {Chen},
  \citenamefont {Wrachtrup}, \citenamefont {Wang},\ and\ \citenamefont
  {Du}}]{Shi.S2015}%
  \BibitemOpen
  \bibfield  {author} {\bibinfo {author} {\bibfnamefont {F.}~\bibnamefont
  {Shi}}, \bibinfo {author} {\bibfnamefont {Q.}~\bibnamefont {Zhang}}, \bibinfo
  {author} {\bibfnamefont {P.}~\bibnamefont {Wang}}, \bibinfo {author}
  {\bibfnamefont {H.}~\bibnamefont {Sun}}, \bibinfo {author} {\bibfnamefont
  {J.}~\bibnamefont {Wang}}, \bibinfo {author} {\bibfnamefont {X.}~\bibnamefont
  {Rong}}, \bibinfo {author} {\bibfnamefont {M.}~\bibnamefont {Chen}}, \bibinfo
  {author} {\bibfnamefont {C.}~\bibnamefont {Ju}}, \bibinfo {author}
  {\bibfnamefont {F.}~\bibnamefont {Reinhard}}, \bibinfo {author}
  {\bibfnamefont {H.}~\bibnamefont {Chen}}, \bibinfo {author} {\bibfnamefont
  {J.}~\bibnamefont {Wrachtrup}}, \bibinfo {author} {\bibfnamefont
  {J.}~\bibnamefont {Wang}}, \ and\ \bibinfo {author} {\bibfnamefont
  {J.}~\bibnamefont {Du}},\ }\bibfield  {title} {\enquote {\bibinfo {title}
  {Single-protein spin resonance spectroscopy under ambient conditions},}\
  }\href {\doibase 10.1126/science.aaa2253} {\bibfield  {journal} {\bibinfo
  {journal} {Science}\ }\textbf {\bibinfo {volume} {347}},\ \bibinfo {pages}
  {1135--1138} (\bibinfo {year} {2015})}\BibitemShut {NoStop}%
\bibitem [{\citenamefont {Rong}\ \emph {et~al.}(2018)\citenamefont {Rong},
  \citenamefont {Wang}, \citenamefont {Geng}, \citenamefont {Qin},
  \citenamefont {Guo}, \citenamefont {Jiao}, \citenamefont {Xie}, \citenamefont
  {Wang}, \citenamefont {Huang}, \citenamefont {Shi}, \citenamefont {Cai},
  \citenamefont {Zou},\ and\ \citenamefont {Du}}]{Rong.nc2018}%
  \BibitemOpen
  \bibfield  {author} {\bibinfo {author} {\bibfnamefont {X.}~\bibnamefont
  {Rong}}, \bibinfo {author} {\bibfnamefont {M.}~\bibnamefont {Wang}}, \bibinfo
  {author} {\bibfnamefont {J.}~\bibnamefont {Geng}}, \bibinfo {author}
  {\bibfnamefont {X.}~\bibnamefont {Qin}}, \bibinfo {author} {\bibfnamefont
  {M.}~\bibnamefont {Guo}}, \bibinfo {author} {\bibfnamefont {M.}~\bibnamefont
  {Jiao}}, \bibinfo {author} {\bibfnamefont {Y.}~\bibnamefont {Xie}}, \bibinfo
  {author} {\bibfnamefont {P.}~\bibnamefont {Wang}}, \bibinfo {author}
  {\bibfnamefont {P.}~\bibnamefont {Huang}}, \bibinfo {author} {\bibfnamefont
  {F.}~\bibnamefont {Shi}}, \bibinfo {author} {\bibfnamefont {Y.-F.}\
  \bibnamefont {Cai}}, \bibinfo {author} {\bibfnamefont {C.}~\bibnamefont
  {Zou}}, \ and\ \bibinfo {author} {\bibfnamefont {J.}~\bibnamefont {Du}},\
  }\bibfield  {title} {\enquote {\bibinfo {title} {Searching for an exotic
  spin-dependent interaction with a single electron-spin quantum sensor},}\
  }\href {\doibase 10.1038/s41467-018-03152-9} {\bibfield  {journal} {\bibinfo
  {journal} {Nature Communications}\ }\textbf {\bibinfo {volume} {9}},\
  \bibinfo {pages} {739} (\bibinfo {year} {2018})}\BibitemShut {NoStop}%
\bibitem [{\citenamefont {Waldherr}\ \emph {et~al.}(2011)\citenamefont
  {Waldherr}, \citenamefont {Neumann}, \citenamefont {Huelga}, \citenamefont
  {Jelezko},\ and\ \citenamefont {Wrachtrup}}]{Waldherr.PRL2011}%
  \BibitemOpen
  \bibfield  {author} {\bibinfo {author} {\bibfnamefont {G.}~\bibnamefont
  {Waldherr}}, \bibinfo {author} {\bibfnamefont {P.}~\bibnamefont {Neumann}},
  \bibinfo {author} {\bibfnamefont {S.~F.}\ \bibnamefont {Huelga}}, \bibinfo
  {author} {\bibfnamefont {F.}~\bibnamefont {Jelezko}}, \ and\ \bibinfo
  {author} {\bibfnamefont {J.}~\bibnamefont {Wrachtrup}},\ }\bibfield  {title}
  {\enquote {\bibinfo {title} {Violation of a temporal \text{B}ell inequality
  for single spins in a diamond defect center},}\ }\href {\doibase
  10.1103/PhysRevLett.107.090401} {\bibfield  {journal} {\bibinfo  {journal}
  {Phys. Rev. Lett.}\ }\textbf {\bibinfo {volume} {107}},\ \bibinfo {pages}
  {090401} (\bibinfo {year} {2011})}\BibitemShut {NoStop}%
\bibitem [{\citenamefont {Hensen}\ \emph {et~al.}(2015)\citenamefont {Hensen},
  \citenamefont {Bernien}, \citenamefont {Dr{\'e}au}, \citenamefont {Reiserer},
  \citenamefont {Kalb}, \citenamefont {Blok}, \citenamefont {Ruitenberg},
  \citenamefont {Vermeulen}, \citenamefont {Schouten}, \citenamefont
  {Abell{\'a}n}, \citenamefont {Amaya}, \citenamefont {Pruneri}, \citenamefont
  {Mitchell}, \citenamefont {Markham}, \citenamefont {Twitchen}, \citenamefont
  {Elkouss}, \citenamefont {Wehner}, \citenamefont {Taminiau},\ and\
  \citenamefont {Hanson}}]{Hensen.N2015}%
  \BibitemOpen
  \bibfield  {author} {\bibinfo {author} {\bibfnamefont {B.}~\bibnamefont
  {Hensen}}, \bibinfo {author} {\bibfnamefont {H.}~\bibnamefont {Bernien}},
  \bibinfo {author} {\bibfnamefont {A.~E.}\ \bibnamefont {Dr{\'e}au}}, \bibinfo
  {author} {\bibfnamefont {A.}~\bibnamefont {Reiserer}}, \bibinfo {author}
  {\bibfnamefont {N.}~\bibnamefont {Kalb}}, \bibinfo {author} {\bibfnamefont
  {M.~S.}\ \bibnamefont {Blok}}, \bibinfo {author} {\bibfnamefont
  {J.}~\bibnamefont {Ruitenberg}}, \bibinfo {author} {\bibfnamefont {R.~F.~L.}\
  \bibnamefont {Vermeulen}}, \bibinfo {author} {\bibfnamefont {R.~N.}\
  \bibnamefont {Schouten}}, \bibinfo {author} {\bibfnamefont {C.}~\bibnamefont
  {Abell{\'a}n}}, \bibinfo {author} {\bibfnamefont {W.}~\bibnamefont {Amaya}},
  \bibinfo {author} {\bibfnamefont {V.}~\bibnamefont {Pruneri}}, \bibinfo
  {author} {\bibfnamefont {M.~W.}\ \bibnamefont {Mitchell}}, \bibinfo {author}
  {\bibfnamefont {M.}~\bibnamefont {Markham}}, \bibinfo {author} {\bibfnamefont
  {D.~J.}\ \bibnamefont {Twitchen}}, \bibinfo {author} {\bibfnamefont
  {D.}~\bibnamefont {Elkouss}}, \bibinfo {author} {\bibfnamefont
  {S.}~\bibnamefont {Wehner}}, \bibinfo {author} {\bibfnamefont {T.~H.}\
  \bibnamefont {Taminiau}}, \ and\ \bibinfo {author} {\bibfnamefont
  {R.}~\bibnamefont {Hanson}},\ }\bibfield  {title} {\enquote {\bibinfo {title}
  {Loophole-free \text{B}ell inequality violation using electron spins
  separated by 1.3 kilometres},}\ }\href {https://doi.org/10.1038/nature15759}
  {\bibfield  {journal} {\bibinfo  {journal} {Nature}\ }\textbf {\bibinfo
  {volume} {526}},\ \bibinfo {pages} {682} (\bibinfo {year}
  {2015})}\BibitemShut {NoStop}%
\bibitem [{\citenamefont {Jin}\ \emph {et~al.}(2017)\citenamefont {Jin},
  \citenamefont {Liu}, \citenamefont {Geng}, \citenamefont {Huang},
  \citenamefont {Ma}, \citenamefont {Shi}, \citenamefont {Duan}, \citenamefont
  {Shi}, \citenamefont {Rong},\ and\ \citenamefont {Du}}]{Jin.PRA2017}%
  \BibitemOpen
  \bibfield  {author} {\bibinfo {author} {\bibfnamefont {F.}~\bibnamefont
  {Jin}}, \bibinfo {author} {\bibfnamefont {Y.}~\bibnamefont {Liu}}, \bibinfo
  {author} {\bibfnamefont {J.}~\bibnamefont {Geng}}, \bibinfo {author}
  {\bibfnamefont {P.}~\bibnamefont {Huang}}, \bibinfo {author} {\bibfnamefont
  {W.}~\bibnamefont {Ma}}, \bibinfo {author} {\bibfnamefont {M.}~\bibnamefont
  {Shi}}, \bibinfo {author} {\bibfnamefont {C.}~\bibnamefont {Duan}}, \bibinfo
  {author} {\bibfnamefont {F.}~\bibnamefont {Shi}}, \bibinfo {author}
  {\bibfnamefont {X.}~\bibnamefont {Rong}}, \ and\ \bibinfo {author}
  {\bibfnamefont {J.}~\bibnamefont {Du}},\ }\bibfield  {title} {\enquote
  {\bibinfo {title} {Experimental test of \text{B}orn's rule by inspecting
  third-order quantum interference on a single spin in solids},}\ }\href
  {\doibase 10.1103/PhysRevA.95.012107} {\bibfield  {journal} {\bibinfo
  {journal} {Phys. Rev. A}\ }\textbf {\bibinfo {volume} {95}},\ \bibinfo
  {pages} {012107} (\bibinfo {year} {2017})}\BibitemShut {NoStop}%
\bibitem [{\citenamefont {Jelezko}\ \emph
  {et~al.}(2004{\natexlab{b}})\citenamefont {Jelezko}, \citenamefont {Gaebel},
  \citenamefont {Popa}, \citenamefont {Domhan}, \citenamefont {Gruber},\ and\
  \citenamefont {Wrachtrup}}]{Jelezko2004}%
  \BibitemOpen
  \bibfield  {author} {\bibinfo {author} {\bibfnamefont {F.}~\bibnamefont
  {Jelezko}}, \bibinfo {author} {\bibfnamefont {T.}~\bibnamefont {Gaebel}},
  \bibinfo {author} {\bibfnamefont {I.}~\bibnamefont {Popa}}, \bibinfo {author}
  {\bibfnamefont {M.}~\bibnamefont {Domhan}}, \bibinfo {author} {\bibfnamefont
  {A.}~\bibnamefont {Gruber}}, \ and\ \bibinfo {author} {\bibfnamefont
  {J.}~\bibnamefont {Wrachtrup}},\ }\bibfield  {title} {\enquote {\bibinfo
  {title} {Observation of coherent oscillation of a single nuclear spin and
  realization of a two-qubit conditional quantum gate},}\ }\href {\doibase
  10.1103/PhysRevLett.93.130501} {\bibfield  {journal} {\bibinfo  {journal}
  {Phys. Rev. Lett.}\ }\textbf {\bibinfo {volume} {93}},\ \bibinfo {pages}
  {130501} (\bibinfo {year} {2004}{\natexlab{b}})}\BibitemShut {NoStop}%
\bibitem [{\citenamefont {Rong}\ \emph {et~al.}(2014)\citenamefont {Rong},
  \citenamefont {Geng}, \citenamefont {Wang}, \citenamefont {Zhang},
  \citenamefont {Ju}, \citenamefont {Shi}, \citenamefont {Duan},\ and\
  \citenamefont {Du}}]{Rong.PRL2014}%
  \BibitemOpen
  \bibfield  {author} {\bibinfo {author} {\bibfnamefont {X.}~\bibnamefont
  {Rong}}, \bibinfo {author} {\bibfnamefont {J.}~\bibnamefont {Geng}}, \bibinfo
  {author} {\bibfnamefont {Z.}~\bibnamefont {Wang}}, \bibinfo {author}
  {\bibfnamefont {Q.}~\bibnamefont {Zhang}}, \bibinfo {author} {\bibfnamefont
  {C.}~\bibnamefont {Ju}}, \bibinfo {author} {\bibfnamefont {F.}~\bibnamefont
  {Shi}}, \bibinfo {author} {\bibfnamefont {C.}~\bibnamefont {Duan}}, \ and\
  \bibinfo {author} {\bibfnamefont {J.}~\bibnamefont {Du}},\ }\bibfield
  {title} {\enquote {\bibinfo {title} {Implementation of dynamically corrected
  gates on a single electron spin in diamond},}\ }\href {\doibase
  10.1103/PhysRevLett.112.050503} {\bibfield  {journal} {\bibinfo  {journal}
  {Phys. Rev. Lett.}\ }\textbf {\bibinfo {volume} {112}},\ \bibinfo {pages}
  {050503} (\bibinfo {year} {2014})}\BibitemShut {NoStop}%
\bibitem [{\citenamefont {Rong}\ \emph {et~al.}(2015)\citenamefont {Rong},
  \citenamefont {Geng}, \citenamefont {Shi}, \citenamefont {Liu}, \citenamefont
  {Xu}, \citenamefont {Ma}, \citenamefont {Kong}, \citenamefont {Jiang},
  \citenamefont {Wu},\ and\ \citenamefont {Du}}]{Rong.nc2015}%
  \BibitemOpen
  \bibfield  {author} {\bibinfo {author} {\bibfnamefont {X.}~\bibnamefont
  {Rong}}, \bibinfo {author} {\bibfnamefont {J.}~\bibnamefont {Geng}}, \bibinfo
  {author} {\bibfnamefont {F.}~\bibnamefont {Shi}}, \bibinfo {author}
  {\bibfnamefont {Y.}~\bibnamefont {Liu}}, \bibinfo {author} {\bibfnamefont
  {K.}~\bibnamefont {Xu}}, \bibinfo {author} {\bibfnamefont {W.}~\bibnamefont
  {Ma}}, \bibinfo {author} {\bibfnamefont {F.}~\bibnamefont {Kong}}, \bibinfo
  {author} {\bibfnamefont {Z.}~\bibnamefont {Jiang}}, \bibinfo {author}
  {\bibfnamefont {Y.}~\bibnamefont {Wu}}, \ and\ \bibinfo {author}
  {\bibfnamefont {J.}~\bibnamefont {Du}},\ }\bibfield  {title} {\enquote
  {\bibinfo {title} {Experimental fault-tolerant universal quantum gates with
  solid-state spins under ambient conditions},}\ }\href {\doibase
  https://doi.org/10.1038/ncomms9748} {\bibfield  {journal} {\bibinfo
  {journal} {Nature Communications}\ }\textbf {\bibinfo {volume} {6}},\
  \bibinfo {pages} {8748} (\bibinfo {year} {2015})}\BibitemShut {NoStop}%
\bibitem [{\citenamefont {Yale}\ \emph {et~al.}(2013)\citenamefont {Yale},
  \citenamefont {Buckley}, \citenamefont {Christle}, \citenamefont {Burkard},
  \citenamefont {Heremans}, \citenamefont {Bassett},\ and\ \citenamefont
  {Awschalom}}]{Christopher.PNAS2013}%
  \BibitemOpen
  \bibfield  {author} {\bibinfo {author} {\bibfnamefont {Christopher~G.}\
  \bibnamefont {Yale}}, \bibinfo {author} {\bibfnamefont {Bob~B.}\ \bibnamefont
  {Buckley}}, \bibinfo {author} {\bibfnamefont {David~J.}\ \bibnamefont
  {Christle}}, \bibinfo {author} {\bibfnamefont {Guido}\ \bibnamefont
  {Burkard}}, \bibinfo {author} {\bibfnamefont {F.~Joseph}\ \bibnamefont
  {Heremans}}, \bibinfo {author} {\bibfnamefont {Lee~C.}\ \bibnamefont
  {Bassett}}, \ and\ \bibinfo {author} {\bibfnamefont {David~D.}\ \bibnamefont
  {Awschalom}},\ }\bibfield  {title} {\enquote {\bibinfo {title} {All-optical
  control of a solid-state spin using coherent dark states},}\ }\href {\doibase
  10.1073/pnas.1305920110} {\bibfield  {journal} {\bibinfo  {journal}
  {Proceedings of the National Academy of Sciences}\ }\textbf {\bibinfo
  {volume} {110}},\ \bibinfo {pages} {7595--7600} (\bibinfo {year}
  {2013})}\BibitemShut {NoStop}%
\bibitem [{\citenamefont {Yale}\ \emph {et~al.}(2016)\citenamefont {Yale},
  \citenamefont {Heremans}, \citenamefont {Zhou}, \citenamefont {Auer},
  \citenamefont {Burkard},\ and\ \citenamefont
  {Awschalom}}]{Christopher.np2016}%
  \BibitemOpen
  \bibfield  {author} {\bibinfo {author} {\bibfnamefont {C.~G.}\ \bibnamefont
  {Yale}}, \bibinfo {author} {\bibfnamefont {F.~J.}\ \bibnamefont {Heremans}},
  \bibinfo {author} {\bibfnamefont {B.~B.}\ \bibnamefont {Zhou}}, \bibinfo
  {author} {\bibfnamefont {A.}~\bibnamefont {Auer}}, \bibinfo {author}
  {\bibfnamefont {G.}~\bibnamefont {Burkard}}, \ and\ \bibinfo {author}
  {\bibfnamefont {D.~D.}\ \bibnamefont {Awschalom}},\ }\bibfield  {title}
  {\enquote {\bibinfo {title} {Optical manipulation of the berry phase in a
  solid-state spin qubit},}\ }\href {https://doi.org/10.1038/nphoton.2015.278}
  {\bibfield  {journal} {\bibinfo  {journal} {Nature Photonics}\ }\textbf
  {\bibinfo {volume} {10}},\ \bibinfo {pages} {184} (\bibinfo {year}
  {2016})}\BibitemShut {NoStop}%
\bibitem [{\citenamefont {Zhou}\ \emph {et~al.}(2016)\citenamefont {Zhou},
  \citenamefont {Baksic}, \citenamefont {Ribeiro}, \citenamefont {Yale},
  \citenamefont {Heremans}, \citenamefont {Jerger}, \citenamefont {Auer},
  \citenamefont {Burkard}, \citenamefont {Clerk},\ and\ \citenamefont
  {Awschalom}}]{Brian.NP2016}%
  \BibitemOpen
  \bibfield  {author} {\bibinfo {author} {\bibfnamefont {Brian~B.}\
  \bibnamefont {Zhou}}, \bibinfo {author} {\bibfnamefont {A.}~\bibnamefont
  {Baksic}}, \bibinfo {author} {\bibfnamefont {H.}~\bibnamefont {Ribeiro}},
  \bibinfo {author} {\bibfnamefont {Christopher~G.}\ \bibnamefont {Yale}},
  \bibinfo {author} {\bibfnamefont {F.~Joseph}\ \bibnamefont {Heremans}},
  \bibinfo {author} {\bibfnamefont {Paul~C.}\ \bibnamefont {Jerger}}, \bibinfo
  {author} {\bibfnamefont {A.}~\bibnamefont {Auer}}, \bibinfo {author}
  {\bibfnamefont {G.}~\bibnamefont {Burkard}}, \bibinfo {author} {\bibfnamefont
  {Aashish~A.}\ \bibnamefont {Clerk}}, \ and\ \bibinfo {author} {\bibfnamefont
  {David~D.}\ \bibnamefont {Awschalom}},\ }\bibfield  {title} {\enquote
  {\bibinfo {title} {Accelerated quantum control using superadiabatic dynamics
  in a solid-state lambda system},}\ }\href {\doibase
  https://doi.org/10.1038/nphys3967} {\bibfield  {journal} {\bibinfo  {journal}
  {Nature Physics}\ }\textbf {\bibinfo {volume} {13}},\ \bibinfo {pages} {330}
  (\bibinfo {year} {2016})}\BibitemShut {NoStop}%
\bibitem [{\citenamefont {Chu}\ and\ \citenamefont
  {Lukin}(2017)}]{Chu.Oxford2015}%
  \BibitemOpen
  \bibfield  {author} {\bibinfo {author} {\bibfnamefont {Y.}~\bibnamefont
  {Chu}}\ and\ \bibinfo {author} {\bibfnamefont {Mikhail~D}\ \bibnamefont
  {Lukin}},\ }\bibfield  {title} {\enquote {\bibinfo {title} {Quantum optics
  with nitrogen-vacancy centers in diamond},}\ }in\ \href@noop {} {\emph
  {\bibinfo {booktitle} {Quantum Optics and Nanophotonics}}}\ (\bibinfo
  {publisher} {Oxford University Press},\ \bibinfo {address} {Oxford},\
  \bibinfo {year} {2017})\BibitemShut {NoStop}%
\bibitem [{\citenamefont {Wang}\ \emph {et~al.}(2014)\citenamefont {Wang},
  \citenamefont {Cai}, \citenamefont {Retzker},\ and\ \citenamefont
  {Plenio}}]{Wang_2014}%
  \BibitemOpen
  \bibfield  {author} {\bibinfo {author} {\bibfnamefont {Z.-Y.}\ \bibnamefont
  {Wang}}, \bibinfo {author} {\bibfnamefont {J.}~\bibnamefont {Cai}}, \bibinfo
  {author} {\bibfnamefont {A.}~\bibnamefont {Retzker}}, \ and\ \bibinfo
  {author} {\bibfnamefont {Martin~B}\ \bibnamefont {Plenio}},\ }\bibfield
  {title} {\enquote {\bibinfo {title} {All-optical magnetic resonance of high
  spectral resolution using a nitrogen-vacancy spin in diamond},}\ }\href
  {\doibase 10.1088/1367-2630/16/8/083033} {\bibfield  {journal} {\bibinfo
  {journal} {New Journal of Physics}\ }\textbf {\bibinfo {volume} {16}},\
  \bibinfo {pages} {083033} (\bibinfo {year} {2014})}\BibitemShut {NoStop}%
\bibitem [{\citenamefont {Bergmann}\ \emph {et~al.}(1998)\citenamefont
  {Bergmann}, \citenamefont {Theuer},\ and\ \citenamefont
  {Shore}}]{Bergmann.RMP1998}%
  \BibitemOpen
  \bibfield  {author} {\bibinfo {author} {\bibfnamefont {K.}~\bibnamefont
  {Bergmann}}, \bibinfo {author} {\bibfnamefont {H.}~\bibnamefont {Theuer}}, \
  and\ \bibinfo {author} {\bibfnamefont {B.~W.}\ \bibnamefont {Shore}},\
  }\bibfield  {title} {\enquote {\bibinfo {title} {Coherent population transfer
  among quantum states of atoms and molecules},}\ }\href {\doibase
  10.1103/RevModPhys.70.1003} {\bibfield  {journal} {\bibinfo  {journal} {Rev.
  Mod. Phys.}\ }\textbf {\bibinfo {volume} {70}},\ \bibinfo {pages}
  {1003--1025} (\bibinfo {year} {1998})}\BibitemShut {NoStop}%
\bibitem [{\citenamefont {Vitanov}\ \emph {et~al.}(2017)\citenamefont
  {Vitanov}, \citenamefont {Rangelov}, \citenamefont {Shore},\ and\
  \citenamefont {Bergmann}}]{Vitanov.RMP2017}%
  \BibitemOpen
  \bibfield  {author} {\bibinfo {author} {\bibfnamefont {Nikolay~V.}\
  \bibnamefont {Vitanov}}, \bibinfo {author} {\bibfnamefont {Andon~A.}\
  \bibnamefont {Rangelov}}, \bibinfo {author} {\bibfnamefont {Bruce~W.}\
  \bibnamefont {Shore}}, \ and\ \bibinfo {author} {\bibfnamefont
  {K.}~\bibnamefont {Bergmann}},\ }\bibfield  {title} {\enquote {\bibinfo
  {title} {Stimulated raman adiabatic passage in physics, chemistry, and
  beyond},}\ }\href {\doibase 10.1103/RevModPhys.89.015006} {\bibfield
  {journal} {\bibinfo  {journal} {Rev. Mod. Phys.}\ }\textbf {\bibinfo {volume}
  {89}},\ \bibinfo {pages} {015006} (\bibinfo {year} {2017})}\BibitemShut
  {NoStop}%
\bibitem [{\citenamefont {Hilser}\ and\ \citenamefont
  {Burkard}(2012)}]{Hilser.PRB2012}%
  \BibitemOpen
  \bibfield  {author} {\bibinfo {author} {\bibfnamefont {F.}~\bibnamefont
  {Hilser}}\ and\ \bibinfo {author} {\bibfnamefont {G.}~\bibnamefont
  {Burkard}},\ }\bibfield  {title} {\enquote {\bibinfo {title} {All-optical
  control of the spin state in the \text{NV} center in diamond},}\ }\href
  {\doibase 10.1103/PhysRevB.86.125204} {\bibfield  {journal} {\bibinfo
  {journal} {Phys. Rev. B}\ }\textbf {\bibinfo {volume} {86}},\ \bibinfo
  {pages} {125204} (\bibinfo {year} {2012})}\BibitemShut {NoStop}%
\bibitem [{\citenamefont {d'Alessandro}(2007)}]{Alessandro.Book2007}%
  \BibitemOpen
  \bibfield  {author} {\bibinfo {author} {\bibfnamefont {D.}~\bibnamefont
  {d'Alessandro}},\ }\href {https://doi.org/10.1201/9781584888833} {\emph
  {\bibinfo {title} {Introduction to quantum control and dynamics}}}\ (\bibinfo
   {publisher} {Chapman and Hall/CRC},\ \bibinfo {year} {2007})\BibitemShut
  {NoStop}%
\bibitem [{\citenamefont {Glaser}\ \emph {et~al.}(2015)\citenamefont {Glaser},
  \citenamefont {Boscain}, \citenamefont {Calarco}, \citenamefont {Koch},
  \citenamefont {Koeckenberger}, \citenamefont {Kosloff}, \citenamefont
  {Kuprov}, \citenamefont {Luy}, \citenamefont {Schirmer}, \citenamefont
  {Schulte-Herbrueggen}, \citenamefont {Sugny},\ and\ \citenamefont
  {Wilhelm}}]{Glaser.EPJD2015}%
  \BibitemOpen
  \bibfield  {author} {\bibinfo {author} {\bibfnamefont {S.~J.}\ \bibnamefont
  {Glaser}}, \bibinfo {author} {\bibfnamefont {U.}~\bibnamefont {Boscain}},
  \bibinfo {author} {\bibfnamefont {T.}~\bibnamefont {Calarco}}, \bibinfo
  {author} {\bibfnamefont {C.~P.}\ \bibnamefont {Koch}}, \bibinfo {author}
  {\bibfnamefont {W.}~\bibnamefont {Koeckenberger}}, \bibinfo {author}
  {\bibfnamefont {R.}~\bibnamefont {Kosloff}}, \bibinfo {author} {\bibfnamefont
  {I.}~\bibnamefont {Kuprov}}, \bibinfo {author} {\bibfnamefont
  {B.}~\bibnamefont {Luy}}, \bibinfo {author} {\bibfnamefont {S.}~\bibnamefont
  {Schirmer}}, \bibinfo {author} {\bibfnamefont {T.}~\bibnamefont
  {Schulte-Herbrueggen}}, \bibinfo {author} {\bibfnamefont {D.}~\bibnamefont
  {Sugny}}, \ and\ \bibinfo {author} {\bibfnamefont {F.~K.}\ \bibnamefont
  {Wilhelm}},\ }\bibfield  {title} {\enquote {\bibinfo {title} {Training
  schr{\"o}dinger's cat: quantum optimal control},}\ }\href {\doibase
  10.1140/epjd/e2015-60464-1} {\bibfield  {journal} {\bibinfo  {journal} {The
  European Physical Journal D}\ }\textbf {\bibinfo {volume} {69}},\ \bibinfo
  {pages} {279} (\bibinfo {year} {2015})}\BibitemShut {NoStop}%
\bibitem [{\citenamefont {Fortunato}\ \emph {et~al.}(2002)\citenamefont
  {Fortunato}, \citenamefont {Pravia}, \citenamefont {Boulant}, \citenamefont
  {Teklemariam}, \citenamefont {Havel},\ and\ \citenamefont
  {Cory}}]{Fortunato.JCP2002}%
  \BibitemOpen
  \bibfield  {author} {\bibinfo {author} {\bibfnamefont {Evan~M.}\ \bibnamefont
  {Fortunato}}, \bibinfo {author} {\bibfnamefont {Marco~A.}\ \bibnamefont
  {Pravia}}, \bibinfo {author} {\bibfnamefont {N.}~\bibnamefont {Boulant}},
  \bibinfo {author} {\bibfnamefont {G.}~\bibnamefont {Teklemariam}}, \bibinfo
  {author} {\bibfnamefont {Timothy~F.}\ \bibnamefont {Havel}}, \ and\ \bibinfo
  {author} {\bibfnamefont {David~G.}\ \bibnamefont {Cory}},\ }\bibfield
  {title} {\enquote {\bibinfo {title} {Design of strongly modulating pulses to
  implement precise effective hamiltonians for quantum information
  processing},}\ }\href {\doibase 10.1063/1.1465412} {\bibfield  {journal}
  {\bibinfo  {journal} {The Journal of Chemical Physics}\ }\textbf {\bibinfo
  {volume} {116}},\ \bibinfo {pages} {7599--7606} (\bibinfo {year}
  {2002})}\BibitemShut {NoStop}%
\bibitem [{\citenamefont {Caneva}\ \emph {et~al.}(2011)\citenamefont {Caneva},
  \citenamefont {Calarco},\ and\ \citenamefont {Montangero}}]{Caneva.PRA2011}%
  \BibitemOpen
  \bibfield  {author} {\bibinfo {author} {\bibfnamefont {T.}~\bibnamefont
  {Caneva}}, \bibinfo {author} {\bibfnamefont {T.}~\bibnamefont {Calarco}}, \
  and\ \bibinfo {author} {\bibfnamefont {S.}~\bibnamefont {Montangero}},\
  }\bibfield  {title} {\enquote {\bibinfo {title} {Chopped random-basis quantum
  optimization},}\ }\href {\doibase 10.1103/PhysRevA.84.022326} {\bibfield
  {journal} {\bibinfo  {journal} {Phys. Rev. A}\ }\textbf {\bibinfo {volume}
  {84}},\ \bibinfo {pages} {022326} (\bibinfo {year} {2011})}\BibitemShut
  {NoStop}%
\bibitem [{\citenamefont {Machnes}\ \emph {et~al.}(2011)\citenamefont
  {Machnes}, \citenamefont {Sander}, \citenamefont {Glaser}, \citenamefont
  {de~Fouquieres}, \citenamefont {Gruslys}, \citenamefont {Schirmer},\ and\
  \citenamefont {Schulte-Herbruggen}}]{Machnes.PRA2011}%
  \BibitemOpen
  \bibfield  {author} {\bibinfo {author} {\bibfnamefont {S.}~\bibnamefont
  {Machnes}}, \bibinfo {author} {\bibfnamefont {U.}~\bibnamefont {Sander}},
  \bibinfo {author} {\bibfnamefont {S.~J.}\ \bibnamefont {Glaser}}, \bibinfo
  {author} {\bibfnamefont {P.}~\bibnamefont {de~Fouquieres}}, \bibinfo {author}
  {\bibfnamefont {A.}~\bibnamefont {Gruslys}}, \bibinfo {author} {\bibfnamefont
  {S.}~\bibnamefont {Schirmer}}, \ and\ \bibinfo {author} {\bibfnamefont
  {T.}~\bibnamefont {Schulte-Herbruggen}},\ }\bibfield  {title} {\enquote
  {\bibinfo {title} {Comparing, optimizing, and benchmarking quantum-control
  algorithms in a unifying programming framework},}\ }\href {\doibase
  10.1103/PhysRevA.84.022305} {\bibfield  {journal} {\bibinfo  {journal} {Phys.
  Rev. A}\ }\textbf {\bibinfo {volume} {84}},\ \bibinfo {pages} {022305}
  (\bibinfo {year} {2011})}\BibitemShut {NoStop}%
\bibitem [{\citenamefont {Ciaramella}\ \emph {et~al.}(2015)\citenamefont
  {Ciaramella}, \citenamefont {Borz}, \citenamefont {Dirr},\ and\ \citenamefont
  {Wachsmuth}}]{Ciaramella.JSC2015}%
  \BibitemOpen
  \bibfield  {author} {\bibinfo {author} {\bibfnamefont {G.}~\bibnamefont
  {Ciaramella}}, \bibinfo {author} {\bibfnamefont {A.}~\bibnamefont {Borz}},
  \bibinfo {author} {\bibfnamefont {G.}~\bibnamefont {Dirr}}, \ and\ \bibinfo
  {author} {\bibfnamefont {D.}~\bibnamefont {Wachsmuth}},\ }\bibfield  {title}
  {\enquote {\bibinfo {title} {Newton methods for the optimal control of closed
  quantum spin systems},}\ }\href {\doibase 10.1137/140966988} {\bibfield
  {journal} {\bibinfo  {journal} {SIAM Journal on Scientific Computing}\
  }\textbf {\bibinfo {volume} {37}},\ \bibinfo {pages} {A319--A346} (\bibinfo
  {year} {2015})}\BibitemShut {NoStop}%
\bibitem [{\citenamefont {Ryan}\ \emph {et~al.}(2010)\citenamefont {Ryan},
  \citenamefont {Hodges},\ and\ \citenamefont {Cory}}]{Ryan.PRL2010}%
  \BibitemOpen
  \bibfield  {author} {\bibinfo {author} {\bibfnamefont {C.~A.}\ \bibnamefont
  {Ryan}}, \bibinfo {author} {\bibfnamefont {J.~S.}\ \bibnamefont {Hodges}}, \
  and\ \bibinfo {author} {\bibfnamefont {D.~G.}\ \bibnamefont {Cory}},\
  }\bibfield  {title} {\enquote {\bibinfo {title} {Robust decoupling techniques
  to extend quantum coherence in diamond},}\ }\href {\doibase
  10.1103/PhysRevLett.105.200402} {\bibfield  {journal} {\bibinfo  {journal}
  {Phys. Rev. Lett.}\ }\textbf {\bibinfo {volume} {105}},\ \bibinfo {pages}
  {200402} (\bibinfo {year} {2010})}\BibitemShut {NoStop}%
\bibitem [{\citenamefont {Machnes}\ \emph {et~al.}(2010)\citenamefont
  {Machnes}, \citenamefont {Plenio}, \citenamefont {Reznik}, \citenamefont
  {Steane},\ and\ \citenamefont {Retzker}}]{Machnes.PRL2010}%
  \BibitemOpen
  \bibfield  {author} {\bibinfo {author} {\bibfnamefont {S.}~\bibnamefont
  {Machnes}}, \bibinfo {author} {\bibfnamefont {M.~B.}\ \bibnamefont {Plenio}},
  \bibinfo {author} {\bibfnamefont {B.}~\bibnamefont {Reznik}}, \bibinfo
  {author} {\bibfnamefont {A.~M.}\ \bibnamefont {Steane}}, \ and\ \bibinfo
  {author} {\bibfnamefont {A.}~\bibnamefont {Retzker}},\ }\bibfield  {title}
  {\enquote {\bibinfo {title} {Superfast laser cooling},}\ }\href {\doibase
  10.1103/PhysRevLett.104.183001} {\bibfield  {journal} {\bibinfo  {journal}
  {Phys. Rev. Lett.}\ }\textbf {\bibinfo {volume} {104}},\ \bibinfo {pages}
  {183001} (\bibinfo {year} {2010})}\BibitemShut {NoStop}%
\bibitem [{\citenamefont {Dolde}\ \emph {et~al.}(2014)\citenamefont {Dolde},
  \citenamefont {Bergholm}, \citenamefont {Wang}, \citenamefont {Jakobi},
  \citenamefont {Naydenov}, \citenamefont {Pezzagna}, \citenamefont {Meijer},
  \citenamefont {Jelezko}, \citenamefont {Neumann}, \citenamefont
  {Schulte-Herbr{\"u}ggen}, \citenamefont {Biamonte},\ and\ \citenamefont
  {Wrachtrup}}]{Dolde.nc2014}%
  \BibitemOpen
  \bibfield  {author} {\bibinfo {author} {\bibfnamefont {F.}~\bibnamefont
  {Dolde}}, \bibinfo {author} {\bibfnamefont {V.}~\bibnamefont {Bergholm}},
  \bibinfo {author} {\bibfnamefont {Y.}~\bibnamefont {Wang}}, \bibinfo {author}
  {\bibfnamefont {I.}~\bibnamefont {Jakobi}}, \bibinfo {author} {\bibfnamefont
  {B.}~\bibnamefont {Naydenov}}, \bibinfo {author} {\bibfnamefont
  {S.}~\bibnamefont {Pezzagna}}, \bibinfo {author} {\bibfnamefont
  {J.}~\bibnamefont {Meijer}}, \bibinfo {author} {\bibfnamefont
  {F.}~\bibnamefont {Jelezko}}, \bibinfo {author} {\bibfnamefont
  {P.}~\bibnamefont {Neumann}}, \bibinfo {author} {\bibfnamefont
  {T.}~\bibnamefont {Schulte-Herbr{\"u}ggen}}, \bibinfo {author} {\bibfnamefont
  {J.}~\bibnamefont {Biamonte}}, \ and\ \bibinfo {author} {\bibfnamefont
  {J.}~\bibnamefont {Wrachtrup}},\ }\bibfield  {title} {\enquote {\bibinfo
  {title} {High-fidelity spin entanglement using optimal control},}\ }\href
  {https://doi.org/10.1038/ncomms4371} {\bibfield  {journal} {\bibinfo
  {journal} {Nature Communications}\ }\textbf {\bibinfo {volume} {5}},\
  \bibinfo {pages} {3371} (\bibinfo {year} {2014})}\BibitemShut {NoStop}%
\bibitem [{\citenamefont {Scheuer}\ \emph {et~al.}(2014)\citenamefont
  {Scheuer}, \citenamefont {Kong}, \citenamefont {Said}, \citenamefont {Chen},
  \citenamefont {Kurz}, \citenamefont {Marseglia}, \citenamefont {Du},
  \citenamefont {Hemmer}, \citenamefont {Montangero}, \citenamefont {Calarco},
  \citenamefont {Naydenov},\ and\ \citenamefont {Jelezko}}]{Scheuer.njp2014}%
  \BibitemOpen
  \bibfield  {author} {\bibinfo {author} {\bibfnamefont {J.}~\bibnamefont
  {Scheuer}}, \bibinfo {author} {\bibfnamefont {X.}~\bibnamefont {Kong}},
  \bibinfo {author} {\bibfnamefont {R.~S.}\ \bibnamefont {Said}}, \bibinfo
  {author} {\bibfnamefont {J.}~\bibnamefont {Chen}}, \bibinfo {author}
  {\bibfnamefont {A.}~\bibnamefont {Kurz}}, \bibinfo {author} {\bibfnamefont
  {L.}~\bibnamefont {Marseglia}}, \bibinfo {author} {\bibfnamefont
  {J.}~\bibnamefont {Du}}, \bibinfo {author} {\bibfnamefont {P.~R}\
  \bibnamefont {Hemmer}}, \bibinfo {author} {\bibfnamefont {S.}~\bibnamefont
  {Montangero}}, \bibinfo {author} {\bibfnamefont {T.}~\bibnamefont {Calarco}},
  \bibinfo {author} {\bibfnamefont {B.}~\bibnamefont {Naydenov}}, \ and\
  \bibinfo {author} {\bibfnamefont {F.}~\bibnamefont {Jelezko}},\ }\bibfield
  {title} {\enquote {\bibinfo {title} {Precise qubit control beyond the
  rotating wave approximation},}\ }\href {\doibase
  10.1088/1367-2630/16/9/093022} {\bibfield  {journal} {\bibinfo  {journal}
  {New Journal of Physics}\ }\textbf {\bibinfo {volume} {16}},\ \bibinfo
  {pages} {093022} (\bibinfo {year} {2014})}\BibitemShut {NoStop}%
\bibitem [{\citenamefont {Geng}\ \emph {et~al.}(2016)\citenamefont {Geng},
  \citenamefont {Wu}, \citenamefont {Wang}, \citenamefont {Xu}, \citenamefont
  {Shi}, \citenamefont {Xie}, \citenamefont {Rong},\ and\ \citenamefont
  {Du}}]{Geng.prl2017}%
  \BibitemOpen
  \bibfield  {author} {\bibinfo {author} {\bibfnamefont {J.}~\bibnamefont
  {Geng}}, \bibinfo {author} {\bibfnamefont {Y.}~\bibnamefont {Wu}}, \bibinfo
  {author} {\bibfnamefont {X.}~\bibnamefont {Wang}}, \bibinfo {author}
  {\bibfnamefont {K.}~\bibnamefont {Xu}}, \bibinfo {author} {\bibfnamefont
  {F.}~\bibnamefont {Shi}}, \bibinfo {author} {\bibfnamefont {Y.}~\bibnamefont
  {Xie}}, \bibinfo {author} {\bibfnamefont {X.}~\bibnamefont {Rong}}, \ and\
  \bibinfo {author} {\bibfnamefont {J.}~\bibnamefont {Du}},\ }\bibfield
  {title} {\enquote {\bibinfo {title} {Experimental time-optimal universal
  control of spin qubits in solids},}\ }\href {\doibase
  10.1103/PhysRevLett.117.170501} {\bibfield  {journal} {\bibinfo  {journal}
  {Phys. Rev. Lett.}\ }\textbf {\bibinfo {volume} {117}},\ \bibinfo {pages}
  {170501} (\bibinfo {year} {2016})}\BibitemShut {NoStop}%
\bibitem [{\citenamefont {Khaneja}\ \emph {et~al.}(2005)\citenamefont
  {Khaneja}, \citenamefont {Reiss}, \citenamefont {Kehlet}, \citenamefont
  {Schulte-Herbrggen},\ and\ \citenamefont {Glaser}}]{Khaneja.JMR2005}%
  \BibitemOpen
  \bibfield  {author} {\bibinfo {author} {\bibfnamefont {N.}~\bibnamefont
  {Khaneja}}, \bibinfo {author} {\bibfnamefont {T.}~\bibnamefont {Reiss}},
  \bibinfo {author} {\bibfnamefont {C.}~\bibnamefont {Kehlet}}, \bibinfo
  {author} {\bibfnamefont {T.}~\bibnamefont {Schulte-Herbrggen}}, \ and\
  \bibinfo {author} {\bibfnamefont {S.~J.}\ \bibnamefont {Glaser}},\ }\bibfield
   {title} {\enquote {\bibinfo {title} {Optimal control of coupled spin
  dynamics: design of \text{NMR} pulse sequences by gradient ascent
  algorithms},}\ }\href {\doibase https://doi.org/10.1016/j.jmr.2004.11.004}
  {\bibfield  {journal} {\bibinfo  {journal} {Journal of Magnetic Resonance}\
  }\textbf {\bibinfo {volume} {172}},\ \bibinfo {pages} {296--305} (\bibinfo
  {year} {2005})}\BibitemShut {NoStop}%
\bibitem [{\citenamefont {To\u{s}ner}\ \emph {et~al.}(2009)\citenamefont
  {To\u{s}ner}, \citenamefont {Vosegaard}, \citenamefont {Kehlet},
  \citenamefont {Khaneja}, \citenamefont {Glaser},\ and\ \citenamefont
  {Nielsen}}]{Tosner.JMR2009}%
  \BibitemOpen
  \bibfield  {author} {\bibinfo {author} {\bibfnamefont {Z.}~\bibnamefont
  {To\u{s}ner}}, \bibinfo {author} {\bibfnamefont {T.}~\bibnamefont
  {Vosegaard}}, \bibinfo {author} {\bibfnamefont {C.}~\bibnamefont {Kehlet}},
  \bibinfo {author} {\bibfnamefont {N.}~\bibnamefont {Khaneja}}, \bibinfo
  {author} {\bibfnamefont {S.~J.}\ \bibnamefont {Glaser}}, \ and\ \bibinfo
  {author} {\bibfnamefont {N.~Chr.}\ \bibnamefont {Nielsen}},\ }\bibfield
  {title} {\enquote {\bibinfo {title} {Optimal control in \text{NMR}
  spectroscopy: Numerical implementation in \text{SIMPSON}},}\ }\href {\doibase
  https://doi.org/10.1016/j.jmr.2008.11.020} {\bibfield  {journal} {\bibinfo
  {journal} {Journal of Magnetic Resonance}\ }\textbf {\bibinfo {volume}
  {197}},\ \bibinfo {pages} {120--134} (\bibinfo {year} {2009})}\BibitemShut
  {NoStop}%
\bibitem [{\citenamefont {Li}\ \emph {et~al.}(2017)\citenamefont {Li},
  \citenamefont {Yang}, \citenamefont {Peng},\ and\ \citenamefont
  {Sun}}]{Li.prl2017}%
  \BibitemOpen
  \bibfield  {author} {\bibinfo {author} {\bibfnamefont {J.}~\bibnamefont
  {Li}}, \bibinfo {author} {\bibfnamefont {X.}~\bibnamefont {Yang}}, \bibinfo
  {author} {\bibfnamefont {X.}~\bibnamefont {Peng}}, \ and\ \bibinfo {author}
  {\bibfnamefont {C.-P.}\ \bibnamefont {Sun}},\ }\bibfield  {title} {\enquote
  {\bibinfo {title} {Hybrid quantum-classical approach to quantum optimal
  control},}\ }\href {\doibase 10.1103/PhysRevLett.118.150503} {\bibfield
  {journal} {\bibinfo  {journal} {Phys. Rev. Lett.}\ }\textbf {\bibinfo
  {volume} {118}},\ \bibinfo {pages} {150503} (\bibinfo {year}
  {2017})}\BibitemShut {NoStop}%
\bibitem [{\citenamefont {Doria}\ \emph {et~al.}(2011)\citenamefont {Doria},
  \citenamefont {Calarco},\ and\ \citenamefont {Montangero}}]{Doria.PRL2011}%
  \BibitemOpen
  \bibfield  {author} {\bibinfo {author} {\bibfnamefont {P.}~\bibnamefont
  {Doria}}, \bibinfo {author} {\bibfnamefont {T.}~\bibnamefont {Calarco}}, \
  and\ \bibinfo {author} {\bibfnamefont {S.}~\bibnamefont {Montangero}},\
  }\bibfield  {title} {\enquote {\bibinfo {title} {Optimal control technique
  for many-body quantum dynamics},}\ }\href {\doibase
  10.1103/PhysRevLett.106.190501} {\bibfield  {journal} {\bibinfo  {journal}
  {Phys. Rev. Lett.}\ }\textbf {\bibinfo {volume} {106}},\ \bibinfo {pages}
  {190501} (\bibinfo {year} {2011})}\BibitemShut {NoStop}%
\bibitem [{\citenamefont {van Frank}\ \emph {et~al.}(2016)\citenamefont {van
  Frank}, \citenamefont {Bonneau}, \citenamefont {Schmiedmayer}, \citenamefont
  {Hild}, \citenamefont {Gross}, \citenamefont {Cheneau}, \citenamefont
  {Bloch}, \citenamefont {Pichler}, \citenamefont {Negretti}, \citenamefont
  {Calarco},\ and\ \citenamefont {Montangero}}]{Frank.SR2016}%
  \BibitemOpen
  \bibfield  {author} {\bibinfo {author} {\bibfnamefont {S.}~\bibnamefont {van
  Frank}}, \bibinfo {author} {\bibfnamefont {M.}~\bibnamefont {Bonneau}},
  \bibinfo {author} {\bibfnamefont {J.}~\bibnamefont {Schmiedmayer}}, \bibinfo
  {author} {\bibfnamefont {S.}~\bibnamefont {Hild}}, \bibinfo {author}
  {\bibfnamefont {C.}~\bibnamefont {Gross}}, \bibinfo {author} {\bibfnamefont
  {M.}~\bibnamefont {Cheneau}}, \bibinfo {author} {\bibfnamefont
  {I.}~\bibnamefont {Bloch}}, \bibinfo {author} {\bibfnamefont
  {T.}~\bibnamefont {Pichler}}, \bibinfo {author} {\bibfnamefont
  {A.}~\bibnamefont {Negretti}}, \bibinfo {author} {\bibfnamefont
  {T.}~\bibnamefont {Calarco}}, \ and\ \bibinfo {author} {\bibfnamefont
  {S.}~\bibnamefont {Montangero}},\ }\bibfield  {title} {\enquote {\bibinfo
  {title} {Optimal control of complex atomic quantum systems},}\ }\href
  {https://doi.org/10.1038/srep34187} {\bibfield  {journal} {\bibinfo
  {journal} {Scientific Reports}\ }\textbf {\bibinfo {volume} {6}},\ \bibinfo
  {pages} {34187} (\bibinfo {year} {2016})}\BibitemShut {NoStop}%
\bibitem [{\citenamefont {Maze}\ \emph {et~al.}(2011)\citenamefont {Maze},
  \citenamefont {Gali}, \citenamefont {Togan}, \citenamefont {Chu},
  \citenamefont {Trifonov}, \citenamefont {Kaxiras},\ and\ \citenamefont
  {Lukin}}]{Maze.njp2011}%
  \BibitemOpen
  \bibfield  {author} {\bibinfo {author} {\bibfnamefont {J.~R.}\ \bibnamefont
  {Maze}}, \bibinfo {author} {\bibfnamefont {A.}~\bibnamefont {Gali}}, \bibinfo
  {author} {\bibfnamefont {E.}~\bibnamefont {Togan}}, \bibinfo {author}
  {\bibfnamefont {Y.}~\bibnamefont {Chu}}, \bibinfo {author} {\bibfnamefont
  {A}~\bibnamefont {Trifonov}}, \bibinfo {author} {\bibfnamefont
  {E.}~\bibnamefont {Kaxiras}}, \ and\ \bibinfo {author} {\bibfnamefont
  {M.~D.}\ \bibnamefont {Lukin}},\ }\bibfield  {title} {\enquote {\bibinfo
  {title} {Properties of nitrogen-vacancy centers in diamond: the group
  theoretic approach},}\ }\href {\doibase 10.1088/1367-2630/13/2/025025}
  {\bibfield  {journal} {\bibinfo  {journal} {New Journal of Physics}\ }\textbf
  {\bibinfo {volume} {13}},\ \bibinfo {pages} {025025} (\bibinfo {year}
  {2011})}\BibitemShut {NoStop}%
\bibitem [{\citenamefont {Manson}\ \emph {et~al.}(2006)\citenamefont {Manson},
  \citenamefont {Harrison},\ and\ \citenamefont {Sellars}}]{Manson.PRB2006}%
  \BibitemOpen
  \bibfield  {author} {\bibinfo {author} {\bibfnamefont {N.~B.}\ \bibnamefont
  {Manson}}, \bibinfo {author} {\bibfnamefont {J.~P.}\ \bibnamefont
  {Harrison}}, \ and\ \bibinfo {author} {\bibfnamefont {M.~J.}\ \bibnamefont
  {Sellars}},\ }\bibfield  {title} {\enquote {\bibinfo {title}
  {Nitrogen-vacancy center in diamond: Model of the electronic structure and
  associated dynamics},}\ }\href {\doibase 10.1103/PhysRevB.74.104303}
  {\bibfield  {journal} {\bibinfo  {journal} {Phys. Rev. B}\ }\textbf {\bibinfo
  {volume} {74}},\ \bibinfo {pages} {104303} (\bibinfo {year}
  {2006})}\BibitemShut {NoStop}%
\bibitem [{\citenamefont {Said}\ and\ \citenamefont
  {Twamley}(2009)}]{Said2009}%
  \BibitemOpen
  \bibfield  {author} {\bibinfo {author} {\bibfnamefont {R.~S.}\ \bibnamefont
  {Said}}\ and\ \bibinfo {author} {\bibfnamefont {J.}~\bibnamefont {Twamley}},\
  }\bibfield  {title} {\enquote {\bibinfo {title} {Robust control of
  entanglement in a nitrogen-vacancy center coupled to a $^{13}\text{C}$
  nuclear spin in diamond},}\ }\href {\doibase 10.1103/PhysRevA.80.032303}
  {\bibfield  {journal} {\bibinfo  {journal} {Phys. Rev. A}\ }\textbf {\bibinfo
  {volume} {80}},\ \bibinfo {pages} {032303} (\bibinfo {year}
  {2009})}\BibitemShut {NoStop}%
\bibitem [{\citenamefont {Bassett}\ \emph {et~al.}(2014)\citenamefont
  {Bassett}, \citenamefont {Heremans}, \citenamefont {Christle}, \citenamefont
  {Yale}, \citenamefont {Burkard}, \citenamefont {Buckley},\ and\ \citenamefont
  {Awschalom}}]{Bassett.S2014}%
  \BibitemOpen
  \bibfield  {author} {\bibinfo {author} {\bibfnamefont {Lee~C.}\ \bibnamefont
  {Bassett}}, \bibinfo {author} {\bibfnamefont {F.~Joseph}\ \bibnamefont
  {Heremans}}, \bibinfo {author} {\bibfnamefont {David~J.}\ \bibnamefont
  {Christle}}, \bibinfo {author} {\bibfnamefont {Christopher~G.}\ \bibnamefont
  {Yale}}, \bibinfo {author} {\bibfnamefont {Guido}\ \bibnamefont {Burkard}},
  \bibinfo {author} {\bibfnamefont {Bob~B.}\ \bibnamefont {Buckley}}, \ and\
  \bibinfo {author} {\bibfnamefont {David~D.}\ \bibnamefont {Awschalom}},\
  }\bibfield  {title} {\enquote {\bibinfo {title} {Ultrafast optical control of
  orbital and spin dynamics in a solid-state defect},}\ }\href {\doibase
  10.1126/science.1255541} {\bibfield  {journal} {\bibinfo  {journal}
  {Science}\ }\textbf {\bibinfo {volume} {345}},\ \bibinfo {pages} {1333--1337}
  (\bibinfo {year} {2014})}\BibitemShut {NoStop}%
\bibitem [{\citenamefont {Fisher}(2010)}]{fisher2010optimal}%
  \BibitemOpen
  \bibfield  {author} {\bibinfo {author} {\bibfnamefont {Robert~M.}\
  \bibnamefont {Fisher}},\ }\emph {\bibinfo {title} {Optimal control of
  multi-level quantum systems}},\ \href
  {https://mediatum.ub.tum.de/node?id=1002028} {Ph.D. thesis},\ \bibinfo
  {school} {Technische Universit{\"a}t M{\"u}nchen} (\bibinfo {year}
  {2010})\BibitemShut {NoStop}%
\end{thebibliography}%

\end{document}